# Peculiarities of the $^4I_{9/2}\rightarrow(^4G_{5/2}+^2G_{7/2})$ transition behavior in $Nd_{0.5}Gd_{0.5}Fe_3(BO_3)_4$ and local magnetic properties of the crystal in the excited states

A V Malakhovskii[1,*], S L Gnatchenko[2], I S Kachur[2], V G Piryatinskaya[2], A L Sukhachev[1] and V L Temerov[1]

[1]L. V. Kirensky Institute of Physics, Siberian Branch of Russian Academy of Sciences, 660036 Krasnoyarsk, Russian Federation

[2]B. Verkin Institute for Low Temperature Physics and Engineering, National Academy of Sciences of Ukraine, 61103 Kharkov, Ukraine

*E-mail adress: malakha@iph.krasn.ru

**Abstract**

Polarized absorption spectra of single-crystal $Nd_{0.5}Gd_{0.5}Fe_3(BO_3)_4$ were studied in the region of transition $^4I_{9/2}\rightarrow(^4G_{5/2}+^2G_{7/2})$ in $Nd^{3+}$ ion as a function of temperature (2 – 34 K) and magnetic field (0 – 65 kOe). Spectra of natural circular dichroism were measured in the range of 5 – 40 K. In the magnetically ordered state, splitting of $Nd^{3+}$ ion excited states due to the exchange interaction of $Nd^{3+}$ and $Fe^{3+}$ ions were determined. It was found out that the local magnetic properties in vicinity of the excited ion substantially depend on the excited state. In particular, a weak ferromagnetic moment appears in some excited states. It was found out that selection rules for transitions between components of the exchange splitting of the ground and excited states substantially deviate from those in paramagnetic state of the crystal. They are different for different transitions and they are very sensitive to orientation of the sublattice magnetic moment relative to the light polarization. In the spectrum of the natural circular dichroism, the transition is revealed which is not observed in the absorption spectrum. The discovered peculiar and multifarious properties of the transitions and states are apparently a consequence of additional quantization axis created by magnetic moments lying in the plane perpendicular to the trigonal axis of the crystal.

FAX: 7-391-2438923, Tel: 7-391-2494556.

## 1. Introduction

Electronically excited atom is, actually, an impurity atom, and, consequently, local properties of crystal in vicinity of the excited atom can change. Spectroscopic manifestations of such local alterations connected with electronic transitions were observed in $RbMnF_3$ and $MnF_2$ [1], in $FeBO_3$ [2] and in some RE containing crystals of huntite structure [3-6]. In [7] it was shown that during the electron transition the initial state of the ion and its interaction with the environment also change and can have an influence upon the polarization of transitions. If there are many excited atoms, not only local properties can change. So, in [8] a phase transition under the influence of the powerful laser pulse was described. Investigation of the local properties of crystals in the optically excited states becomes important in the recent years in connection with the problem of the quantum information processing (see e. g., [9-12]). Crystals containing RE ions are widely used in these efforts. For example, a change of the local crystal properties near the optically excited atom was used for read out of information in the quantum memory [10]. The present work is devoted to study of some *f-f* electron transitions in the single-crystal $Nd_{0.5}Gd_{0.5}Fe_3(BO_3)_4$, which revealed unusual polarization properties. Additionally, behavior of the transitions as a function of temperature and magnetic field allowed us to make conclusions about local magnetic properties of the crystal in the region of the optically excited atom.

A number of RE ferroborates, refer to multiferroics [13-17], i. e., they possess magnetic and electric order simultaneously. $Nd_{0.5}Gd_{0.5}Fe_3(BO_3)_4$ is also the multiferroic [18]. All RE ferroborates are magnetically ordered at temperatures below 30 - 40 K. The crystal $Nd_{0.5}Gd_{0.5}Fe_3(BO_3)_4$ is easy plane antiferromagnet from $T_N$ = 32 K down to at least 2 K [19]. For future discussion it is important that at temperatures $T$ < 11 K it was found a hysteresis in magnetization of the crystal in the easy plane, indicating appearance of the static magnetic domains [19]. The crystal has trigonal symmetry with the space group $R32$ and the lattice constants are: $a$= 9.557(7) Å and $c$= 7.62(1) Å [19]. Trivalent RE ions occupy $D_3$ symmetry positions. They are located at the center of trigonal prisms made up of six crystallography equivalent oxygen ions. The triangles formed by the oxygen ions in the neighboring basal planes are not superimposed on each other but are twisted through a particular angle. The $FeO_6$ octahedrons share edges in such a way that they form helicoidal chains, which run parallel to the $C_3$ axis and are mutually independent. All Fe ions occupy $C_2$-symmetry positions. Structural phase transitions were not found down to 2 K [19]. Magnetic structure of the $Nd_{0.5}Gd_{0.5}Fe_3(BO_3)_4$ crystal was not studied in detail. However, its magnetic properties [19] are close to those of the related crystal $NdFe_3(BO_3)_4$ [20-23]. Therefore, it is possible to suppose that magnetic structure of these crystals is also similar. In particular, neutron diffraction measurements of $NdFe_3(^{11}BO_3)_4$ testified to magnetic spiral configurations with the magnetic moments oriented parallel to the hexagonal basal plane [24]. Later [25] it was shown that in the commensurate magnetic phase below $T_N \approx 30$ K all three magnetic Fe moments and the magnetic Nd moment are aligned ferromagnetically in the basal hexagonal plane but align antiferromagnetically between adjacent planes. It was also shown that in the incommensurate spiral magnetic phase (below $T \approx 13.5$ K) the magnetic structure of $NdFe_3(^{11}BO_3)_4$ is transformed into a long-period antiferromagnetic helix with single chirality. In Ref. [26] it was shown that this phase transition behaves as the first order one. Nonresonant x-ray magnetic scattering has shown that the correlation length (or size) of the magnetic domains is around 100 Å [27]. An element selective resonant magnetic x-ray scattering study has confirmed that the magnetic order of the Nd sublattice is induced by the Fe spin order [28]. When a magnetic field is applied parallel to the hexagonal basal plane, the helicoidal spin order is suppressed and a collinear ordering, where the moments are forced to align in a direction perpendicular to the applied magnetic field, is stabilized [28].

Absorption spectra of $Nd_{0.5}Gd_{0.5}Fe_3(BO_3)_4$ single crystal were earlier analyzed with the help of the Judd-Ofelt theory and spectroscopic characteristics of the crystal were obtained [29]. Optical and magneto-optical properties of the $Nd_{0.5}Gd_{0.5}Fe_3(BO_3)_4$ crystal in the near IR spectral region were studied in [30]. Optical spectra and crystal field parameters of the related crystal $NdFe_3(BO_3)_4$ were studied in [23].

## 2. Experimental details

$Nd_{0.5}Gd_{0.5}Fe_3(BO_3)_4$ single crystals were grown from the melt solution on the base of $K_2Mo_3O_{10}$ as described in [31]. The sample used for optical absorption measurements was 0.2 mm-thick plane-parallel polished plate oriented parallel to the crystallographic axis $C_3$. Absorption spectra were measured using diffraction monochromator MDR-23 with diffraction grating 1200 lines/mm and linear dispersion 1.3 nm/mm. The spectral resolution was about 1.5 $cm^{-1}$ in the studied spectral region. The light intensity was measured by photomultiplier with further computer registration. The absorption spectra were measured with the light propagating normal to the $C_3$ axis of the crystal, electric vector of light being parallel (the π-spectrum) or perpendicular (the σ-spectrum) to the $C_3$ axis. The light was polarized by the Glan prism.

Natural circular dichroism (NCD) spectra were studied on the sample of 0.215 mm-thick cut perpendicular to the $C_3$ axis of the crystal and with light propagated parallel to $C_3$ axis (α-polarization). The NCD spectra were measured by the method of light-polarization modulation using a piezoelectric modulator (details see in [32]). Spectral resolution at NCD measurements was about 2.3 $cm^{-1}$.

Magnetic field was created by a superconducting solenoid with Helmholtz type coils. The magnetic field direction was parallel to the surface of the sample perpendicular or parallel to the $C_3$ axis (Fig. 1). The superconducting solenoid with the sample was placed in liquid helium and all measurements in magnetic field were fulfilled at $T = 2$ K. For the temperature measurements of absorption and natural circular dichroism spectra a liquid-helium cooled cryostat was used. It had an internal volume filled by gaseous helium where the sample was placed. The temperature of the sample was regulated by heating element.

## 3. Results and discussion

### 3.1. Identification of excited states

Polarized absorption spectra of the $Nd_{0.5}Gd_{0.5}Fe_3(BO_3)_4$ single crystal in the region of transition ${}^4I_{9/2}\rightarrow({}^4G_{5/2}+{}^2G_{7/2})$ at $T$=6 K and at $T$=33 K (above $T_N$) are shown in Figs. 2a and 2b respectively. Symmetry of the ground state $Gr1$ of $Nd^{3+}$ ion in the crystal was identified earlier [30] (see Table 1). Excited state of the $D$-manifold (${}^4G_{5/2}+{}^2G_{7/2}$) is split in the crystal field of $D_3$ symmetry in the following way: ${}^4G_{5/2}$: $2E_{1/2}+E_{3/2}$ and ${}^2G_{7/2}$: $3E_{1/2}+E_{3/2}$. Symmetries of states in the $D$-manifold are found (Table 1) according to linear polarizations of the absorption lines (Fig. 2a, Table 1), selection rules of Table 2 and symmetry of the ground state. Polarization of $D1(Gr2)$ and $D2(Gr2)$ transitions (Fig. 2b) from the first excited state $Gr2$ of the ground manifold gives symmetry of this state (Table 1). Shape of $D1(Gr2)$ and $D2(Gr2)$ lines is apparently due to phonon side-bands caused by acoustic phonons.

In a trigonal crystal, for half integer total moment there are three possible values of the crystal quantum number [33]: $\mu$ = +1/2, -1/2, 3/2 (±3/2). States with $M_J = \mu \pm 3n$ (where $n$=0, 1, 2, …) correspond to each $\mu$ in the trigonal symmetry [33]. As a result, the following set of states is obtained:

$$M_J = \pm 1/2,\ \pm 3/2,\ \pm 5/2,\ \pm 7/2,\ \pm 9/2;$$
$$\mu = \pm 1/2, (\pm 3/2), \mp 1/2, \pm 1/2, (\pm 3/2). \qquad (1)$$

The set of the crystal field states for $J$-multiplets of $Nd^{3+}$ ion is evidently found according to value of $J$. States with $\mu = \pm 1/2$ correspond to states $E_{1/2}$ and states with $\mu = (\pm 3/2)$ correspond to states $E_{3/2}$ in the $D_3$ group notations.

Electron states of free atom in a homogeneous electric field ($C_{\infty v}$ symmetry) are split according to the absolute value of the magnetic quantum number $M_J$. Electron states of atom in the trigonal crystal field are also split according to absolute values of $M_J$, in a first approximation. Therefore, the atom wave functions can be described by $|J,\pm M_J\rangle$ states. In this approximation, the effective Landé factor of the Kramers duplets along $C_3$ axis is defined by the equation [7]:

$$g_{CM} = 2gM_J \qquad (2)$$

where $g$ is the Landé factor of the free atom. Results for $Nd^{3+}$ ion are shown in Table 3. States with the same $\mu$ and different $M_J$ (see Eq. 1) can mix in the crystal, and resulting $g_C$ can be both smaller and larger than $g_{CM}$. The prevailing $M_J$ states of the free atom in the crystal field states of the $D$-manifold (Table 1) can be found basing on the comparison of $g_{CM}$ for the corresponding $M_J$ (Tables 3 and 1) with the theoretical $g_C$ in the $NdFe_3(BO_3)_4$ crystal (Table 1). They are of course different but succession of values permits to identify origin of $D$-states from the $M_J$ states (see Table 1).

Presented selection rules and identifications refer to paramagnetic state of the crystal. Magnetic ordering introduces substantial changes in the situation.

### 3.2. Behavior of absorption lines as a function of magnetic field and temperature

#### D1 line

At temperatures $T<T_N$ electronic states of $Nd^{3+}$ are split by the exchange field of magnetically ordered Fe-subsystem (Fig. 3). As mentioned above, the measurements in magnetic field were fulfilled at $T$=2 K. At this temperature only transitions from the lower sublevel of the ground state exchange splitting can be observed. (Generally, there are possible four transitions between components of the ground and excited state exchange splitting (Fig. 3)). At the zero magnetic field, $D1\sigma$ line consists of two components of the Gaussian shape: $D1a\sigma$ and $D1c\sigma$ (see Fig. 4), corresponding to transitions into components of the excited state exchange splitting (Fig. 3). We supposed that the more intensive component, $D1a\sigma$, corresponds to transition without overturn of the Nd ion magnetic moment (Fig. 3). Positions of $D1a\sigma$ and $D1c\sigma$ lines as a function of magnetic field directed perpendicular to $C_3$ axis are shown in Fig. 5. In the same figure there is the field dependence of the $D1\pi$ line position. $D1\pi$ line is not decomposed on components and can be identified as $D1a\pi$ line. So, $D1c$ line is not active in $\pi$-polarization (see Fig. 3). In Fig. 5 (inset) there is the field dependence of position of the undecomposed $D1\sigma$ line maximum, which demonstrates hysteresis in the crystal remagnetization [19]. Fig. 6 shows variation of the lines intensities as a function of magnetic field perpendicular to $C_3$ axis. From Fig. 6 it follows that in the field $H$>4 kOe the $D1c\sigma$ line disappears. $H$=4 kOe is the spin-flop and simultaneously one domain state field in the basal plane [19] (see also Fig. 5, inset). In this case magnetic moments in all domains become perpendicular to the magnetic field and to the $\sigma$-polarization (Fig. 1). Thus, $D1c$ line is allowed only in $\sigma$-polarization and when $M||\sigma$ (Fig. 3). At higher fields a component of magnetic moment $M||\sigma$ appears and, correspondingly, $D1c\sigma$ line appears again (Fig. 6). Intensity of $D1a\pi$ line practically does not depend on magnetic field (Fig. 6) since always $\pi\perp M$ (Fig. 1). Orientation of magnetic moments depending on magnetic field $H\perp C_3$ in the region of 0-4 kOe does not influence the intensity of $D1a\sigma$ line (Fig. 6) but influences its energy (Fig. 5). It is a consequence of different energy of different domains in magnetic field. The energy of $D1a\pi$ line also decreases in the same field region (Fig. 5) but in a less degree, i. e., it seems that the energy of the state depends on the transition polarization. But apparently, we deal with two absorbing objects: domains and domain walls with different energies and polarizations of the transitions. Indeed, at $H$>4 kOe, when there are no domain walls, energies of $D1a\sigma$ and $D1a\pi$ lines coincide. $D1$ lines are good approximated by Gaussians that testifies to the inhomogeneous nature of the line widths. The $D1a\sigma$ line width decreases in the region of 0-4 kOe (Fig. 6, inset), when the crystal transfers to the one domain state. This corresponds to transfer to higher magnetic homogeneity of the crystal. $D1a\pi$ line does not reveal such behavior. This is a consequence of the mentioned above smaller influence of magnetic state on the $D1a\pi$ transition energy.

After spin-flop, energies of both exchange split components of antiferromagnet change identically in the magnetic field due to sublattice angularity caused by the magnetic field. So, the changing of a line frequency, defined by the difference of the ground and excited state energies variations, should be identical for lines $D1a\sigma$ and $D1c\sigma$, corresponding to transitions into sublevels of the excited state exchange splitting. This contradicts to Fig. 5. Observed dependences of Fig. 5 can be accounted for in assumption that spontaneous sublattice angularity (weak ferromagnetic moment $\Delta M$) occurs in the excited $D1$ state. This ferromagnetic moment has opposite direction in $D1^{(+)}$ and $D1^{(-)}$ states (Fig. 3). Energy of this moment in the magnetic field changes also in opposite directions in $D1^{(+)}$ and $D1^{(-)}$ states. In magnetic field parallel to $C_3$ axis energies of the considered absorption lines steadily increase, i. e., the spontaneous sublattice angularity does not occur in this direction.

With the increasing temperature, transitions $D1b$ and $D1d$ (Fig. 3) from the upper sublevel of the ground state exchange splitting should appear on the lower energy side of the absorption spectrum. The ground state exchange splitting at 6 K is ~9 cm$^{-1}$ [30]. Therefore, energy of $D1b$ transition at 6 K should be 9 cm$^{-1}$ less than that of $D1a$ transition and so D1b transition is certainly not observed in absorption spectra of Fig. 7, inset. Absorption spectra of $D1$ transition in σ-polarization (Fig. 7, inset) were decomposed on Gaussian components $D1a\sigma$ and $D1c\sigma$ in the temperature range 6-33 K, and temperature dependences of positions and intensities of the components were obtained (Figs. 7 and 8). Temperature behavior of energy of the $\pi$-polarized $D1a$ line ($D1a\pi$ line) is also shown in Fig. 7. Difference of the $D1a\sigma$ and $D1c\sigma$ lines energies at $T$=6-16 K (Fig. 7) gives exchange splitting of the excited state of ~8 $cm^{-1}$ (Fig. 3). The energy of $D1d$ transition is close to that of $D1a$ transition (Fig. 3), since the exchange splitting of the excited state at 6 K is close to that of the ground state and therefore $D1d$ transition can take part in $D1a$ line intensity at increasing temperature. $D1$ transition in paramagnetic state of the crystal is active both in $\pi$- and $\sigma$-polarizations (Table 1). However, in magnetically ordered state transition $D1c$ is active only in $\sigma$-polarization and only when $M||\sigma$.

From Fig. 8, it is seen that at $T$~20 K $D1c\sigma$ line disappears. Observed phenomenon is evidently connected with the exchange interaction in the excited state and has a threshold character. This can be explained by appearance of the mentioned above weak ferromagnetic moment $\Delta M$ in the excited state at $T$<20 K. It is quite natural that the ferromagnetic moment appears at $T<T_N$. Figs. 7 and 8 testify that the splitting between $D1a\sigma$ and $D1c\sigma$ lines is defined by the antiferromagnetic exchange interaction, but intensity of $D1c\sigma$ line increases with increasing weak ferromagnetic moment $\Delta M$. Thus, it would be more properly to characterize selection rule for $D1c\sigma$ line not relative to sublattice magnetic moment $M$ but relative to ferromagnetic moment: $\sigma\perp\Delta M$.

**D2 line**

Fig. 9 presents $\pi$-spectra of $D2$ line at three temperatures. Spectra of σ-polarized $D2$ line at three temperatures are shown in Fig. 9, inset. At $T$=2 K only transitions from the lower sublevel of the ground state exchange splitting are observed. $D2\pi$ and $D2\sigma$ lines at this temperature are not split (are good approximated by one Gauss function) and have different energy. Therefore, they can be considered as separate transitions to sublevels of the excited state exchange splitting (see the $D2$ transitions diagram in Fig. 10). At the increasing temperature, the $D2b\pi$ line (Fig. 9) corresponding to the transition from the upper sublevel of the ground state exchange splitting appears. Positions of lines $D2a\pi$ and $D2b\pi$ as a function of temperature (Fig. 11) were found as positions of negative extremums in the second derivatives spectra. $D2d\sigma$ line (Fig. 9, inset), corresponding to transition between the upper sublevels of the ground and excited states exchange splitting (Fig. 10), is very weak, and its dependence on temperature can not be found. Position of $D2c\sigma$ line (Fig. 10) is given in Fig. 11 as position of maximum of the total σ-spectrum. At low temperature, when only the lower sublevel of the ground state exchange splitting is occupied, and at temperature near $T_N$ this curve gives correct position of $D2c\sigma$ line. So, distance between $D2a\pi$ and $D2c\sigma$ lines gives exchange splitting of the excited state at 6 K equal to ~2 $cm^{-1}$. In temperature behavior of $D2$ lines (Fig. 11) there are no indications of any features in the region of $T$=20 K, unlike $D1$ line. The Landé factor $g_{\perp}$ along magnetic moments is not equal to zero in the $D2$ state (see Table 1). Therefore, both the ground and the excited states should take part in the exchange splitting between $D2a\pi$ and $D2b\pi$ lines. Exchange splitting of $\pi$-polarized line, found from Fig. 11 (~14 $cm^{-1}$ at 6 K), is larger than the found earlier [30] exchange splitting of the ground state (~9 $cm^{-1}$), i.e., exchange splitting of the ground and excited states are summarized. This is possible in the case of the transitions diagram shown in Fig. 10. Sum of the excited state exchange splitting and found earlier splitting of the ground state is not exactly equal to the splitting between $D2a\pi$ and $D2b\pi$ lines. However, it is necessary to note that the ground state exchange splitting found from different transitions are also appreciably different [30].

There are two variants of the transitions diagram from the view-point of the energetically favorable orientation of the sublattice magnetic moment in the excited state. In the first one it is the same as that in the ground state and then the transitions $D2a$ and $D2b$ occur with overturn of the magnetic moment direction. In the second variant everything is vice-versa (Fig. 10). We have chosen the second variant, although there is no yet a strong criterion for the choice. In this case, transitions $D2c$ and $D2d$ occur with the change of the magnetic moment direction.

At $T$=2 K of magnetic measurements actually only transitions $D2a\pi$ and $D2c\sigma$ (Fig. 10) from the lower component of the ground state exchange splitting are observed (Fig. 9). Variations of positions and intensities of these lines as a function of magnetic field $H\perp C_3$ are presented in Figs. 12 and 12, inset, respectively. As mentioned above, at $H$=4 kOe magnetic moments of all domains are perpendicular to $\sigma$-polarization (Fig. 1). Then, from the dependence of the intensity in the field range 0-4 kOe (Fig. 12, inset) it follows that, in contrast to $D1$ transition, $D2c\sigma$ transition is more intensive for $M\perp\sigma$ than for $M||\sigma$. Behavior of the $D2a\pi$ and $D2c\sigma$ line positions in magnetic field $H\perp C_3$ (Fig. 12) is qualitatively the same as that of $D1a\pi$ and $D1c\sigma$ lines (Fig. 5) and can be explained similarly. In particular, in the region of 0-4 kOe the change of the lines energies are due to different energy of different domains in magnetic field. The same as in the case of $D1$ line, opposite change of the $D2a\pi$ and $D2c\sigma$ transitions energy in magnetic field H>4 kOe (Fig. 12) can be accounted for in assumption that spontaneous sublattice angularity occurs (weak ferromagnetic moment $\Delta M$) in the excited $D2$ state. In magnetic field parallel to $C_3$ axis energies of the considered absorption lines steadily increase. The same as in the case of $D1$ state, this means that the spontaneous sublattice angularity does not occur in this direction.

**D4 line**

Absorption spectrum of the $D4$ transition in σ-polarization at 2 K and $H$=0 is similar to that of $D1$ line (Fig. 4). The spectrum is decomposed on two components of the Gauss shape, which can be identified as transitions from the lower component of the ground state exchange splitting to components of the exchange splitting of the excited state (Fig. 13). We supposed that the more strong transition $D4a\sigma$ occurs without overturn of the ion magnetic moment. In this case energetically favorable orientation of the sublattice magnetic moment in the excited state is the same as in the ground state (Fig. 13). Energies of transitions $D4a\sigma$ and $D4c\sigma$ change identically in magnetic field $H$>4 kOe (Fig. 14). Consequently, in the $D4$ excited state magnetic moments are oriented purely antiferromagnetically (Fig. 13). Absorption spectrum of the $D4$ line in $\pi$-polarization is not decomposed on components. Energy of this line in the zero magnetic field coincides within the limit of experimental error with that of the $D4a\sigma$ line (Fig. 14). Thus, $\pi$-polarized $D4$ line can be identified as the $D4a\pi$ line, and the $D4c$ line is not active in $\pi$-polarization (Fig. 13). Unlike the $D1$ line, behavior of the $D4a\pi$ and $D4a\sigma$ lines energies at $H$<4 kOe is similar, but dependence of the $D4a\pi$ line position on the magnetic field at $H$>4 kOe qualitatively differs from that of the $D4a\sigma$ line (Fig. 14). In the noncentro-symmetrical crystal $Nd_{0.5}Gd_{0.5}Fe_3(BO_3)_4$, inversion twins can exist. They can be the objects with different optical properties in the magnetic field, which correspond to two field dependencies in two polarizations. Positions of $D4a\sigma$ and $D4a\pi$ lines in magnetic field $H||C_3$ change qualitatively similarly to those in the field $H\perp C_3$ (Fig. 14) i.e., also differently. Magnetic field dependence of the $D4a\sigma$ line intensity (Fig. 14, inset) shows that absorption probability for $M\perp\sigma$ prevails in this transition.

Transformation of the $D4\sigma$ line with temperature could not be studied because of influence of adjacent absorption lines (Figs. 2a and 2b). Absorption spectrum of the $D4\pi$ line is not split with the temperature increasing from 2 to 33 K. Consequently, $D4b$ and $D4d$ lines (Fig. 13), possible with increasing temperature, are not observed, i. e., these transitions are forbidden (at least in $\pi$-polarization). Diagrams of $D4$ (Fig. 13) and $D1$ (Fig. 3) transitions are similar to some extent. However, intensity of the $D4c\sigma$ transition is only slightly sensitive to orientation of the domains magnetic moments relative to $\sigma$-polarization, while $D1c\sigma$ transition is allowed only for $\sigma||M$. Additionally, in $D4$ state there is no indications to existence of weak ferromagnetic moment.

**D5 line**

At temperature 2 K, when only lower ground state sublevel is occupied, the $D5\pi$ line is not split (Fig. 15, inset). Consequently, only transition to one component of the excited state exchange splitting is observed. At higher temperatures the $D5\pi$ line is split on two temperature dependent components: $D5a\pi$ and $D5b\pi$ (Fig. 15, inset). Temperature dependence of their positions is depicted in Fig. 15. Splitting of 16.5 $cm^{-1}$ between these lines at 8 K is substantially larger than the exchange splitting of the ground state. Consequently, exchange splitting of the ground and excited states are summarized and the *D*5 transitions diagram should have the form, shown in Fig. 16. If we suppose that transition *D5a* occurs without overturn of the ion magnetic moment, then it inevitably follows that energetically favorable orientation of the ion magnetic moment in the excited state is opposite to that in the ground state (Fig. 16). Position of the *D5aπ* line as a function of magnetic field $H \perp C_3$ at $T$=2 K is shown in Fig. 17.

The *D5σ* line at $T$=2 K is decomposed on two components (Fig. 18, inset), which can be identified as *D5aσ* and *D5cσ* lines (see diagram of Fig. 16). Position of the *D5aσ* line as a function of magnetic field $H \perp C_3$ coincides with that of the *D5aπ* line at $H$>4 kOe (Fig. 17), but differs from it at $H$<4 kOe. This is a feature already mentioned above in the *D*1 line. Dependences of the *D5aσ* and *D5cσ* lines energies on magnetic field $H \perp C_3$ are parallel within the limit of the experimental error (Fig. 17), i. e. orientation of magnetic moments in the *D*5 state are purely antiferromagnetic ones in magnetic field $H \perp C_3$. Intensities of *D5aσ* and *D5cσ* lines are very sensitive to orientation of the domains magnetic moments relative to σ-polarization (Fig. 18): the *D5aσ* line has mainly $\sigma \perp M$ polarization while the *D5cσ* line has mainly $\sigma || M$ polarization, but not purely $\sigma || M$ polarization as it was in the *D*1*c* transition (Fig. 3).

The *D*5 transition reveals peculiar behavior in magnetic field $H || C_3$. The *D5π* line is not split in $H \perp C_3$ field. However in the field $H || C_3$ a splitting appears (Fig. 19, inset). From the field dependence of intensities of the splitting components (Fig. 19) it is seen that the new absorption line appears only at $H$>30 kOe. The appeared line is identified as *D5cπ* line according to its energy (see Fig. 17). The *D5aπ* and *D5cπ* lines energies as a function of $H || C_3$ are also presented in Fig. 17. Opposite direction of these dependences allows us to assume, that magnetic field $H || C_3$ stimulates appearance of a weak ferromagnetic moment $\Delta M$, whose direction in $D5^{(-)}$ state coincides with the field direction. The threshold field of the weak ferromagnetic moment inducing is not known exactly since dependence of the *D5cπ* line intensity on the $\Delta M$ is not known. Therefore $H$=30 kOe of the additional line appearance can be considered only as the approximate threshold field value.

**D6 line**

This absorption line, both in $\pi$ and $\sigma$ polarizations, is very nice approximated by the Lorentz curve and is not split with the increasing temperature and magnetic field. At $T$=2 K there is a small splitting (~0.4 cm$^{-1}$) between $\pi$ and $\sigma$ polarized lines. It would be possible to refer this splitting to experimental error, but field dependences of the lines positions (Fig. 20) show that these are two transitions into sublevels of the excited state exchange splitting (Fig. 21). Variations of *D6aσ* and *D6cπ* lines positions in magnetic field $H \perp C_3$ (Fig. 20) show that a weak spontaneous ferromagnetic moment appears in the *D*6 excited state. Diagram of Fig. 21 was drawn with this circumstance taking into account. The exchange splitting of the $D6$ state is very small. Therefore influence of the external magnetic field is relatively large. There is a substantial asymmetry of the field dependencies (Fig. 20) since the external magnetic field also induces angularity of magnetic moments, but of the same sign in both excited states. Fig. 20, inset testifies that in the magnetic field $H||C_3$ a weak ferromagnetic moment also exists, but magnetic states are inverted. The weak ferromagnetic moments, the most probably, are stimulated by the magnetic field. Dependence of the $D6a\sigma$ line intensity on magnetic field $H\perp C_3$ (Fig. 22) shows that polarization $M||\sigma$ prevails in this absorption.

Excited states $D3$ and $D7$ have the Landé factor $g_{\perp}=0$ (see Table 1). Therefore, the splitting of the absorption lines in the exchange field of the iron ordered in the plain $\perp C_3$ should be equal to the exchange splitting of the ground state. However such splitting are not observed. This means that transitions from the upper sublevel of the ground state exchange splitting are forbidden in $D3$ and $D7$ transitions.

### 3.3. Natural circular dichroism

Natural circular dichroism (NCD) can exist if a crystal structure has no centre of inversion, and NCD was really observed in the studied crystal. Presence of NCD means also that inverse twins of one type prevail. NCD is measured in α-polarized light. NCD spectra of the $D$-band at several temperatures are depicted in Fig. 23. Only $D1$ line demonstrates substantial changes with the temperature decrease and it was studied in detail. Transformation of the NCD spectrum of the $D1$ line with temperature is shown in Fig. 24 and temperature dependences of the line positions are presented in Fig. 25. The substantial temperature dependent line splitting is observed. For electric dipole transitions $\alpha$-polarization is equivalent to $\sigma$-polarization. However, no any splitting connected with the exchange interaction in the ground state was observed both in σ and π absorption spectra of the $D1$ line (Fig. 7), while additional temperature dependent absorption line $D1b$ (Fig. 3) could appear in this case from the low energy side of lines $D1a\sigma$ and $D1a\pi$. According above consideration this means that some transitions forbidden in absorption are allowed in NCD.

Natural optical activity (NOA) is defined by the formula:

$$a_N = \frac{R_{if}}{D_{if}} \approx \frac{\langle \Delta k \rangle_0}{\langle k \rangle_0}, \tag{3}$$

where $k$ is coefficient of absorption and $\Delta k$ is NCD. According to [34]:

$$R_{if} = \mathrm{Im}\left[\langle i|\vec{d}|f\rangle\langle f|\vec{m}|i\rangle\right], \tag{4}$$

where $\vec{d}$ and $\vec{m}$ are electric and magnetic dipole moments, respectively. $D_{if} = \left|\langle i|\vec{d}|f\rangle\right|^2$, since considered transitions are mainly of the electric dipole nature. Due to presence of the magnetic dipole matrix element in (4), selection rules for the NCD can differ from those for electric dipole transitions and other transitions can be allowed in the NCD. In particular, two lines observed in the NCD spectra apparently correspond to *D*1*c* and *D*1*b* transitions (see Fig. 3). Indeed, energy *E*2 in the region of 10-15 K (Figs. 24 and 25) coincides with the energy of the D1cσ absorption line (Fig. 7). Maximum splitting between *D*1*a* and *D*1*c* lines (Fig. 3) is ~8 $cm^{-1}$ (Fig. 7) and together with the exchange splitting of the ground state ~9 $cm^{-1}$ it gives splitting 17 $cm^{-1}$ between *D*1*c* and *D*1*b* lines. This value is close to maximum splitting of ~20 $cm^{-1}$ in the NCD spectrum (Fig. 25). Additionally, sharp decrease of the splitting in the NCD spectrum at 20 K correlates with disappearance of the *D*1*cσ* transition (Figs. 7 and 8). Fig. 25 allows us to suppose, that at *T*=20 K the *D*1*b* line in NCD also disappears (in absorption it was not observed at all) and only the *D*1*a* line remains. Similar situation takes place at 5 K. Thus, at *T*<5 K and *T*>20 K only the *D*1*a* line remains in the NCD spectrum (Figs 24 and 25). The *D*1*b* and *D*1*c* transitions have one common feature: they occur with the overturn of the ion magnetic moment (Fig. 3). Disappearance of the *D*1*b* and *D*1*c* lines in NCD at *T*=5 K permits us to infer that at this temperature there are some additional local transformations which are not revealed in absorption spectra.

## 4. Summary

Polarized absorption spectra of the $Nd_{0.5}Gd_{0.5}Fe_3(BO_3)_4$ crystal in the region of transition $^4I_{9/2}\rightarrow(^4G_{5/2}+^2G_{7/2})$ in $Nd^{3+}$ ion were studied as a function of temperature (2 – 34 K) and magnetic field (0 – 65 kOe). Spectra of natural circular dichroism were measured in the range of 5 – 40 K. Symmetries of $Nd^{3+}$ ion states in the crystal field of $D_3$ local symmetry were identified in paramagnetic state of the crystal. In magnetically ordered state, the splitting of $Nd^{3+}$ ion excited states due to the exchange interaction of $Nd^{3+}$ and $Fe^{3+}$ ions were determined. Results of investigations in the magnetically ordered state revealed four sets of phenomena.

1. Local magnetic properties in vicinity of the excited ion substantially depend on the excited state. In particular, values of the $Nd^{3+}$ ion exchange splitting in excited states are different (table 1) and in a number of the excited states (*D*1, *D*2 and *D*6) a weak ferromagnetic moment appears (Figs. 3, 10 and 21). In the *D*5 state a weak ferromagnetic moment is stimulated by the external magnetic field in the $C_3$ direction. In states *D*2 and *D*5 (Figs. 10 and 16) energetically favorable orientation of the $Nd^{3+}$ ion magnetic moment is opposite to that in the ground state (see also [35]). These observations refer to the more general problem of light induced magnetic phenomena [36].

2. Selection rules for transitions between components of the exchange splitting of the ground and excited states substantially deviate from those in paramagnetic state of the crystal. They are different for different transitions in spite of the identical symmetry of the studied excited states in the local $D_3$ symmetry of the crystal in the ground state (Table 1). This testifies to distortions of the local symmetry in the excited states. In the spectrum of the natural circular dichroism the transition was revealed that was not observed in the absorption spectrum. This also testifies to a specific selection rules in the magnetically ordered state.

3. Intensities of transitions substantially depend on orientation of the sublattice magnetic moment relative to the light polarization, but these dependences are qualitatively different for different transitions. For example, the $D1c$ transition (Fig. 3) is allowed only in σ-polarization and only when $M||\sigma$; the $D5c$ transition (Fig. 16) appears only when the magnetic moment $\Delta M||C_3$ appears in magnetic field and so on. It is necessary to remind that in paramagnetic state all discussed transitions are allowed both in π and σ polarizations (Table 1).

4. In some cases, energies of transitions depend on polarization. In particular, in the region 0-4 kOe this occurs in $D1$ and $D5$ transitions (Figs 5 and 17). At $H$>4 kOe (one domain state) this occurs in $D4$ transition (Fig. 14). Such phenomena can be accounted for by absorption in two objects. In the former case these can be domains and domain walls, in the latter case – inversion twins.

It is evident that the magnetic ordering of the crystal is the source of described peculiar and multifarious properties of selection rules, transitions and electron states. In particular, it is apparently important that magnetic moments in the studied crystal are in plane perpendicular to $C_3$ axis (Fig. 1). Therefore, they create second quantization axis besides $C_3$, and, as a consequence, they create specific selection rules for electron transitions, which depend also on the excited states since electronically excited atom changes the local symmetry and magnetic properties of the crystal. All considered states have $E_{1/2}$ symmetry in $D_3$ local crystal symmetry (Table 1). However they have different values of $J$ or $M_J$ or both of them (Table 1) and, therefore, interactions of the atom with the environment are different in these states and the local properties of the crystal are also different.

**Acknowledgements**

The work was supported by the Russian Foundation for Basic Researches Grant 12-02-00026 and by the President of Russia grant No Nsh-2886.2014.2.

**Figure captions**

Fig. 1. Geometry of experiments.
Fig. 2a. Polarized absorption spectra of $^4I_{9/2}\rightarrow(^4G_{5/2}+^2G_{7/2})$ transition (*D*-band) at 6 K.
Fig. 2b. Polarized absorption spectra of $^4I_{9/2}\rightarrow(^4G_{5/2}+^2G_{7/2})$ transition (*D*-band) at 33 K.
Fig. 3. Diagram of the *D*1 transitions.
Fig. 4. σ-polarized absorption spectra of *D*1 line at $T$=2 K in magnetic field $H\perp C_3$.
Fig. 5. The *D*1 transitions energies as a function of magnetic field $H\perp C_3$ at $T$=2 K. Inset: position of the undecomposed $D1\sigma$ line maximum.
Fig. 6. The *D*1 transitions intensities as a function of magnetic field $H\perp C_3$ at $T$=2 K. Inset: line widths as a function of magnetic field $H\perp C_3$.
Fig. 7. The *D*1 transitions energies as a function of temperature in zero magnetic field. Inset: σ-polarized absorption lines at several temperatures.
Fig. 8. The *D*1 transitions intensities as a function of temperature in zero magnetic field.
Fig. 9. $\pi$-polarized absorption spectra of the *D*2 transitions at several temperatures in zero magnetic field. Inset: the same at $\sigma$-polarization.
Fig. 10. Diagram of the *D*2 transitions.
Fig. 11. The *D*2 transitions energies as a function of temperature in zero magnetic field.
Fig. 12. The *D*2 transitions energies as a function of magnetic field $H\perp C_3$ at $T$=2 K. Inset: the *D*2 transitions intensities as a function of magnetic field $H\perp C_3$ at $T$=2 K.
Fig. 13. Diagram of the *D*4 transitions.
Fig. 14. The *D*4 transitions energies as a function of magnetic field $H\perp C_3$ and $H\|C_3$ at $T$=2 K. Inset: the *D*4 transitions intensities as a function of magnetic field $H\perp C_3$ at $T$=2 K.
Fig. 15. The *D*5 transitions energies as a function of temperature in zero magnetic field. Inset: $\pi$-polarized absorption lines at several temperatures.
Fig. 16. Diagram of the *D*5 transitions.
Fig. 17. The *D*5 transitions energies as a function of magnetic field $H\perp C_3$ and $H\|C_3$ at $T$=2 K.
Fig. 18. The *D*5 transitions intensities as a function of magnetic field $H\perp C_3$ at $T$=2 K. Inset: $\sigma$-polarized absorption spectra of the *D*5 transition at two magnetic fields.

Fig. 19. The $D5$ transitions intensities as a function of magnetic field $H||C_3$ at $T$=2 K. Inset: $\pi$-polarized absorption spectra of the $D5$ transition at several magnetic fields.

Fig. 20. The $D6$ transitions energies as a function of magnetic field $H\perp C_3$ and $H||C_3$ at $T$=2 K.

Fig. 21. Diagram of the $D6$ transitions.

Fig. 22. The $D6$ transitions intensities as a function of magnetic field $H\perp C_3$ at $T$=2 K.

Fig. 23. NCD spectra of the $^4I_{9/2}\rightarrow(^4G_{5/2}+^2G_{7/2})$ transition ($D$-band).

Fig. 24. NCD spectra in the region of the $D1$ line.

Fig. 25. Energies of components of the NCD spectrum in the region of the $D1$ line as a function of temperature.

Table 1. Parameters of transitions and states. ΔE1 – exchange splitting of absorption lines, connected with the exchange splitting of the ground state, ΔE2 - exchange splitting of excited states, $g_{CM}$ - values of $g_C$ in approximation of $|J,\pm M_J\rangle$ states of the free atom. Energies of transitions (E) are given at 40 K. (Details are in the text).

| State | Level | E $cm^{-1}$ | Polar. | Sym. | μ | M | ΔE1 (π) $cm^{-1}$ | ΔE1 (σ) $cm^{-1}$ | ΔE2 $cm^{-1}$ | $g_\perp$ [23] | $g_C$ [23] | $g_{CM}$ |
|---|---|---|---|---|---|---|---|---|---|---|---|---|
| $^4I_{9/2}$ | Gr1 | 0 | - | $E_{1/2}$ | ∓ 1/2 | ±5/2 | | | | 2.385 | 1.376 | 3.64 |
| | Gr2 | ~78 | - | $E_{3/2}$ | 3/2 | ±9/2 | | | | 0 | 3.947 | 6.54 |
| | | | | | | | | | | | | |
| $^4G_{5/2}$ | D1 | 16921 | π, σ | $E_{1/2}$ | ±1/2 | ±1/2 | | | 7 | 0.043 | 0.065 | 0.571 |
| | D2 | 17062 | π, σ | $E_{1/2}$ | ∓ 1/2 | ±5/2 | 14.2 | | 2 | 1.385 | 1.310 | 2.885 |
| | D3 | 17100 | ≈ σ | $E_{3/2}$ | 3/2 | ±3/2 | | | 0 | 0 | 3.044 | 1.713 |
| $^2G_{7/2}$ | D4 | 17199 | π, σ | $E_{1/2}$ | ±1/2 | ±1/2 | | | 12 | 2.617 | 0.266 | 0.889 |
| | D5 | 17240 | π, σ | $E_{1/2}$ | ±1/2 | ±7/2 | ~17 | | 7.5 | 0.755 | 3.308 | 6.223 |
| | D6 | 17289 | π, σ | $E_{1/2}$ | ∓ 1/2 | ±5/2 | | | 0.4 | 1.538 | 0.954 | 4.445 |
| | D7 | 17325 | ≈ σ | $E_{3/2}$ | 3/2 | ±3/2 | | | 0 | 0 | 1.016 | 2.667 |

Table 2. Selection rules for electric dipole transitions in $D_3$ symmetry

| | $E_{1/2}$ | $E_{3/2}$ |
|---|---|---|
| $E_{1/2}$ | π, σ(α) | σ(α) |
| $E_{3/2}$ | σ(α) | π |

Table 3. Landé factors of the Kramers doublets along $C_3$ axis of a crystal in approximation of $|J,\pm M_J\rangle$ states of the free atom.

| | *M* | 13/2 | 11/2 | 9/2 | 7/2 | 5/2 | 3/2 | 1/2 |
|---|---|---|---|---|---|---|---|---|
| $^4I_{9/2}$, g=0.727 | $g_{Cmax}$ | | | 6.54 | 5.09 | 3.64 | 2.18 | 0.727 |
| $^4G_{5/2}$, g=0.571 | $g_{Cmax}$ | | | | | 2.855 | 1.713 | 0.571 |
| $^2G_{7/2}$, g=0.889 | $g_{Cmax}$ | | | | 6.223 | 4.445 | 2.667 | 0.889 |

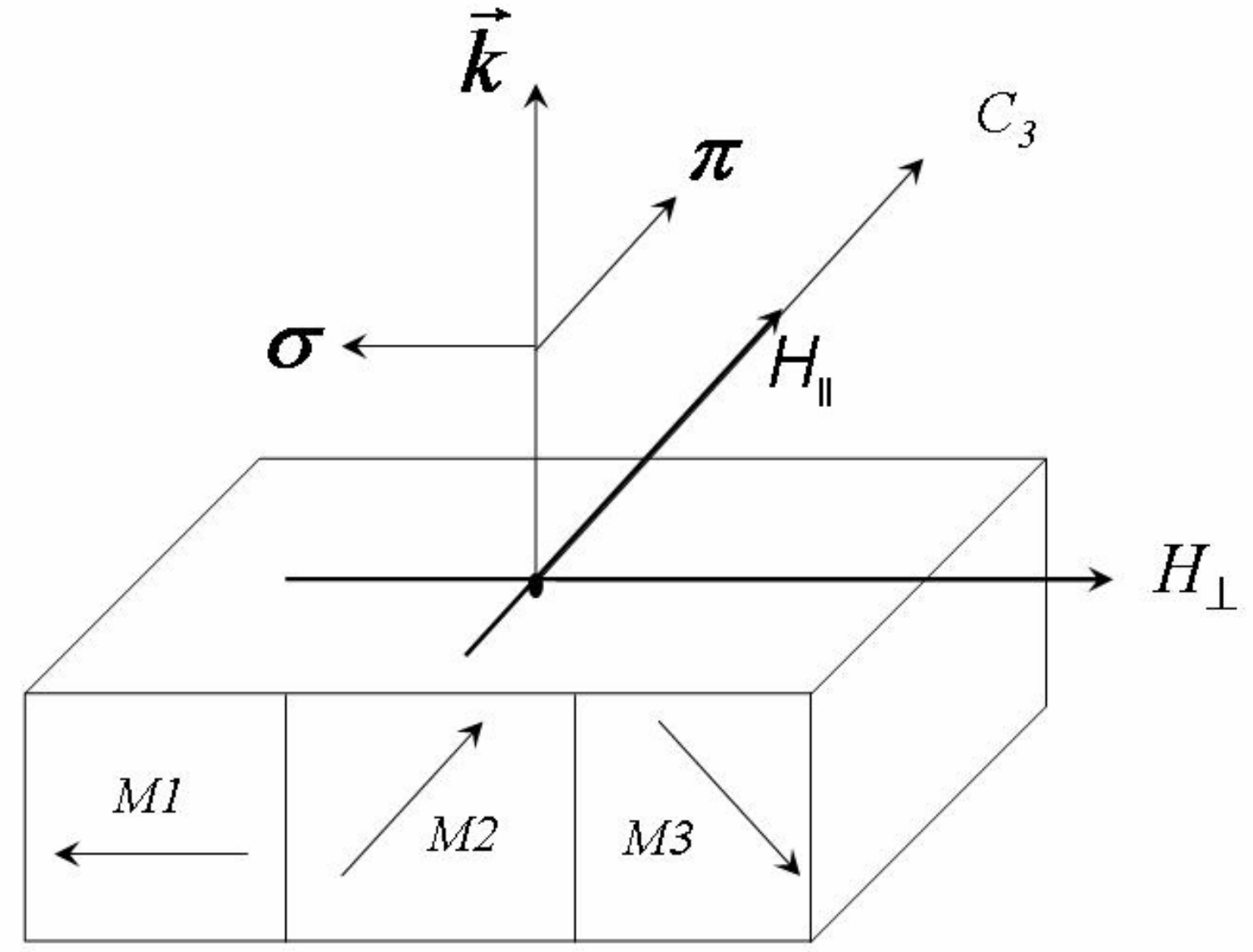

Fig. 1. Geometry of experiments.

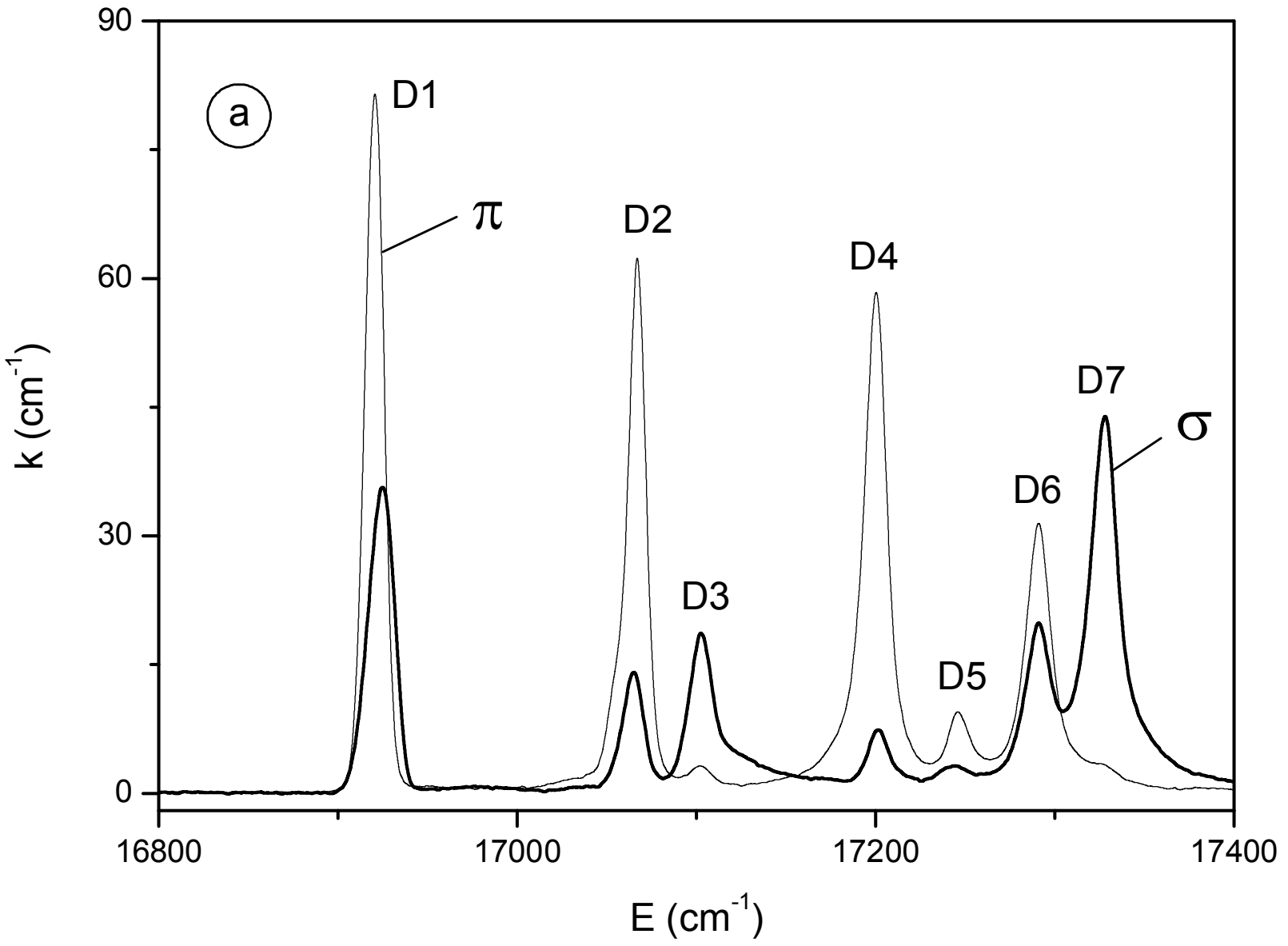

Fig. 2a. Polarized absorption spectra of $^4I_{9/2}\rightarrow(^4G_{5/2}+^2G_{7/2})$ transition (D-band) at 6 K.

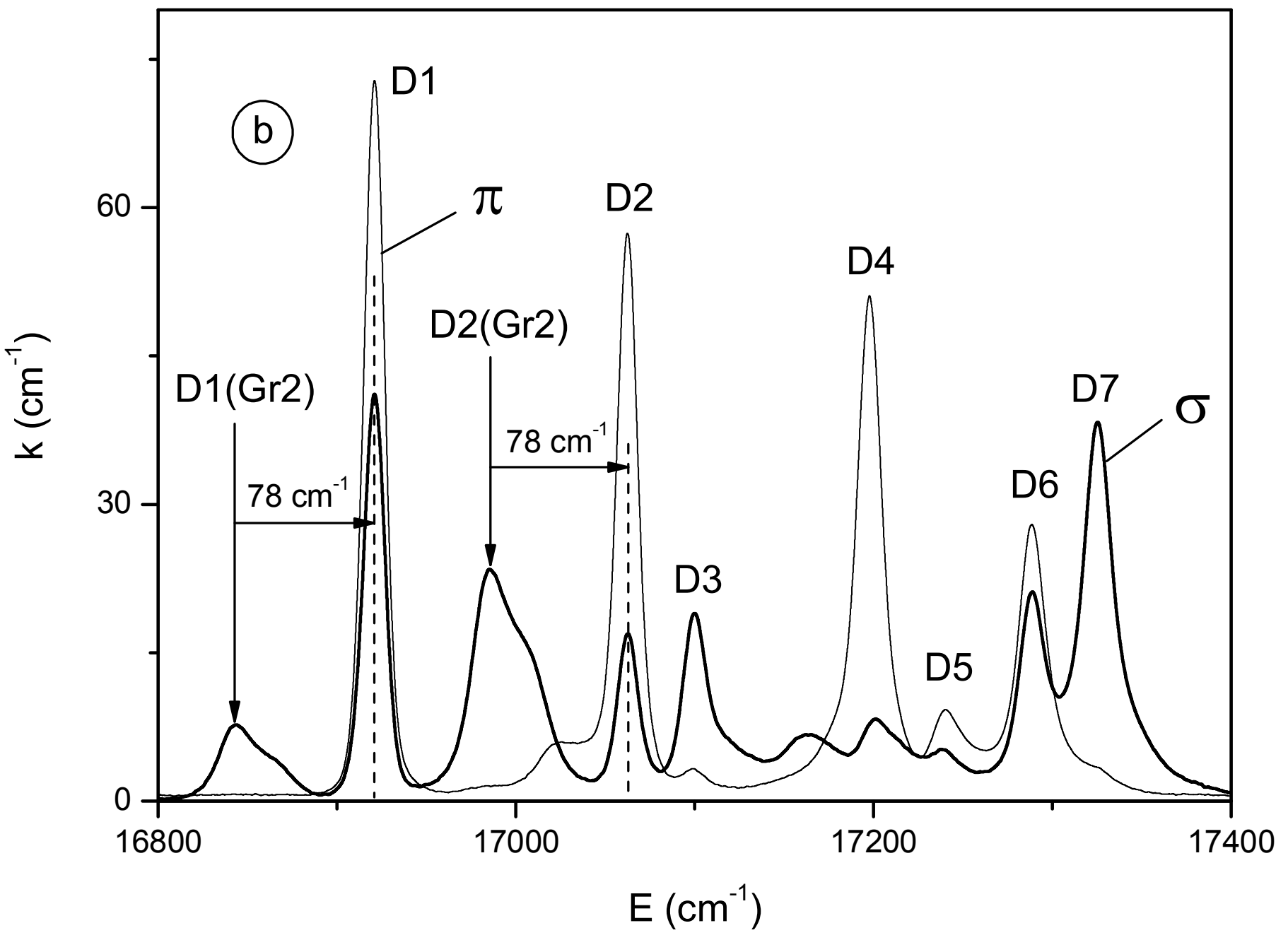

Fig. 2b. Polarized absorption spectra of $^4I_{9/2}\rightarrow(^4G_{5/2}+^2G_{7/2})$ transition (D-band) at 33 K.

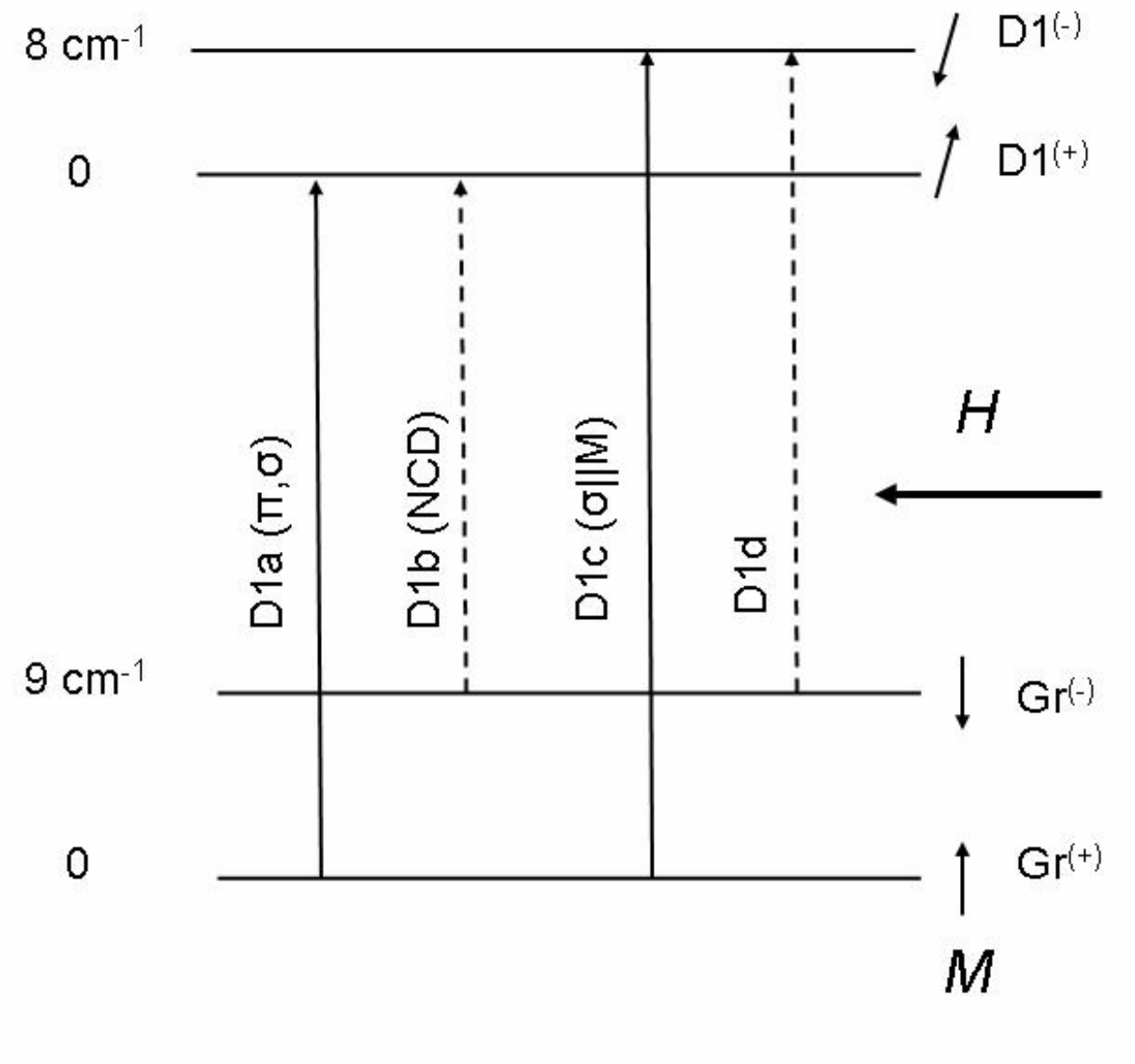

Fig. 3. Diagram of the D1 transition.

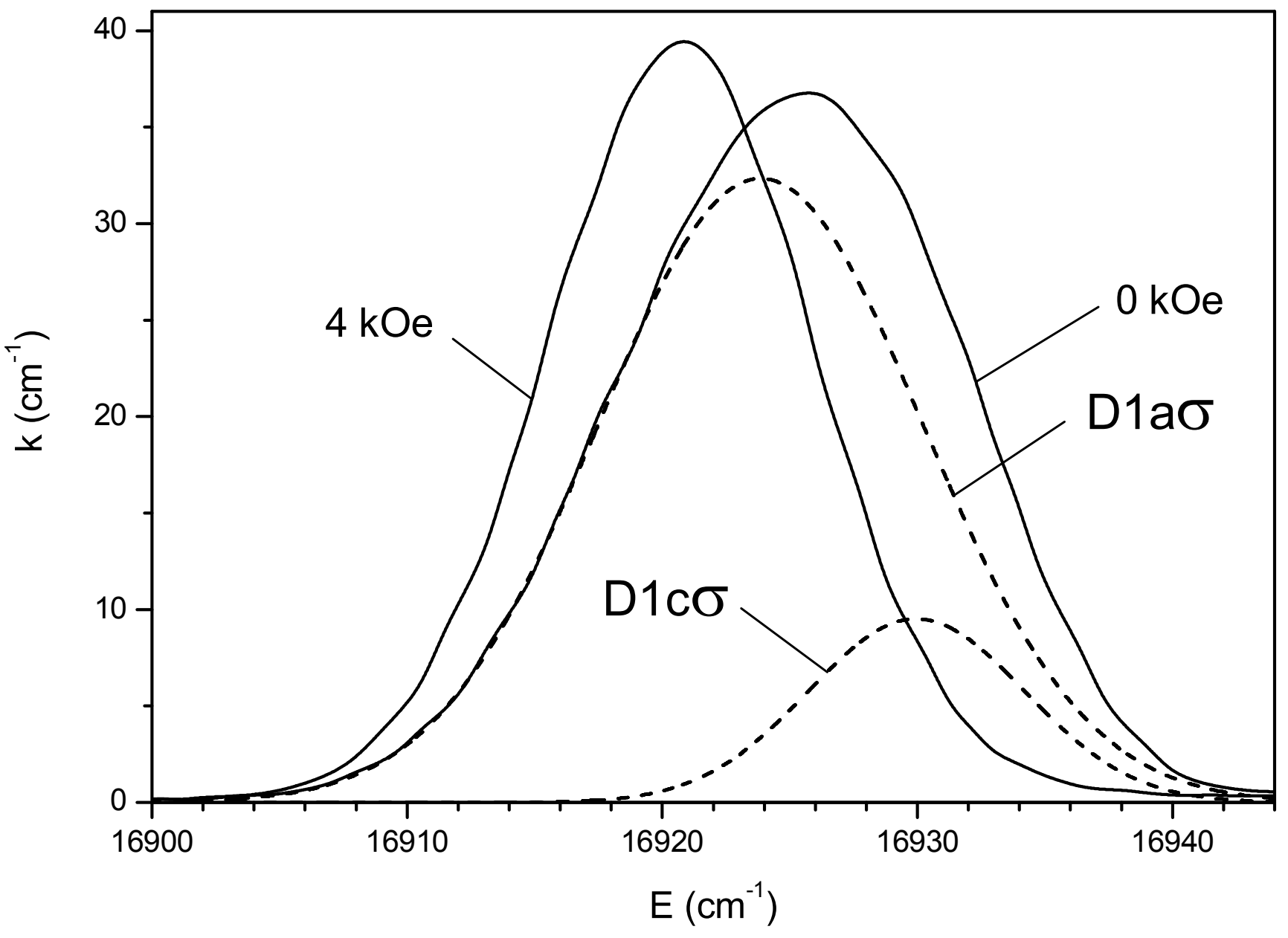

Fig. 4. σ-polarized absorption spectra of D1 line at T=2 K in magnetic field H⊥$C_3$.

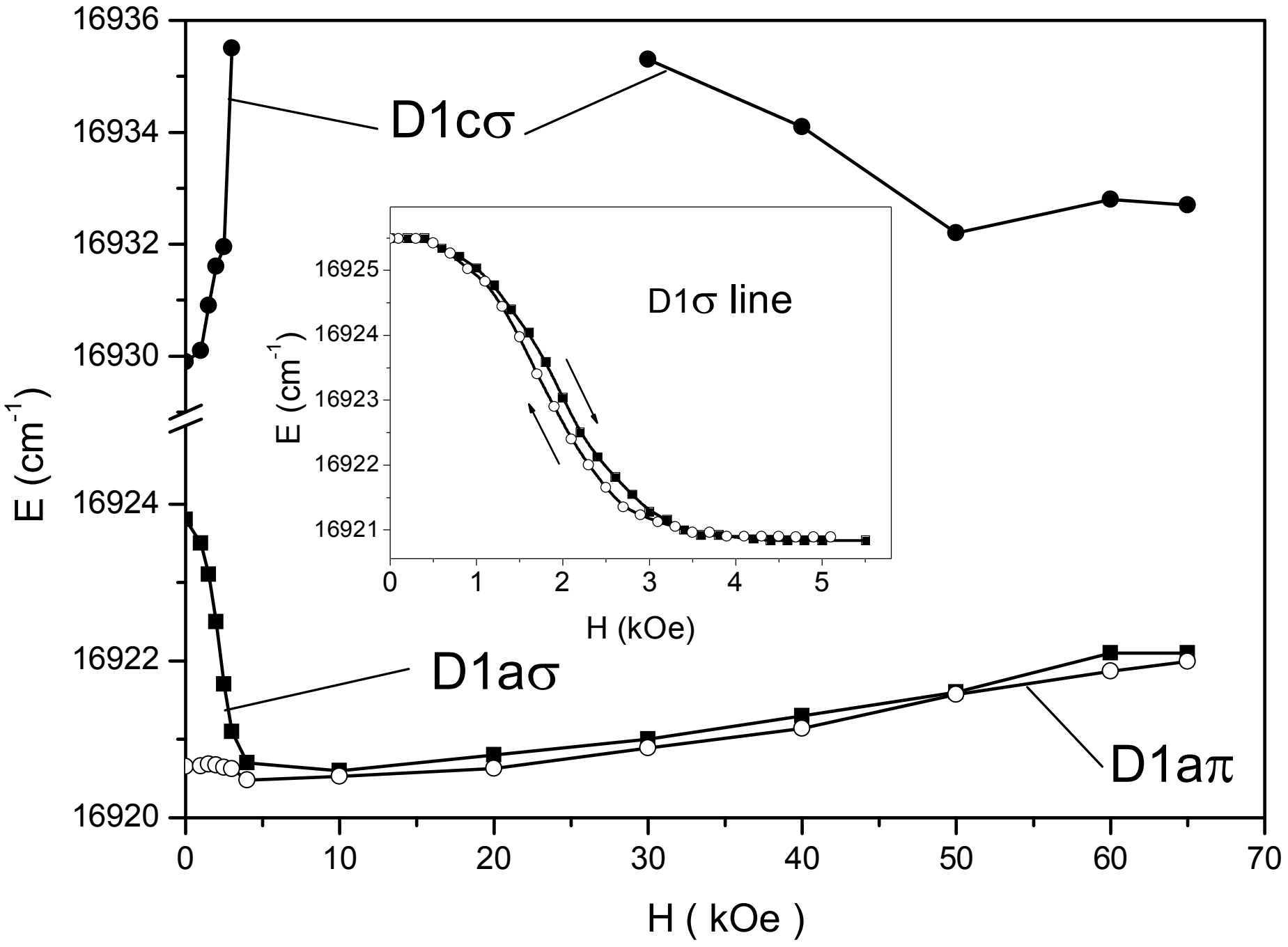

Fig. 5. The D1 transitions energies as a function of magnetic field H⊥$C_3$ at T=2 K. Inset: position of the undecomposed D1σ line maximum.

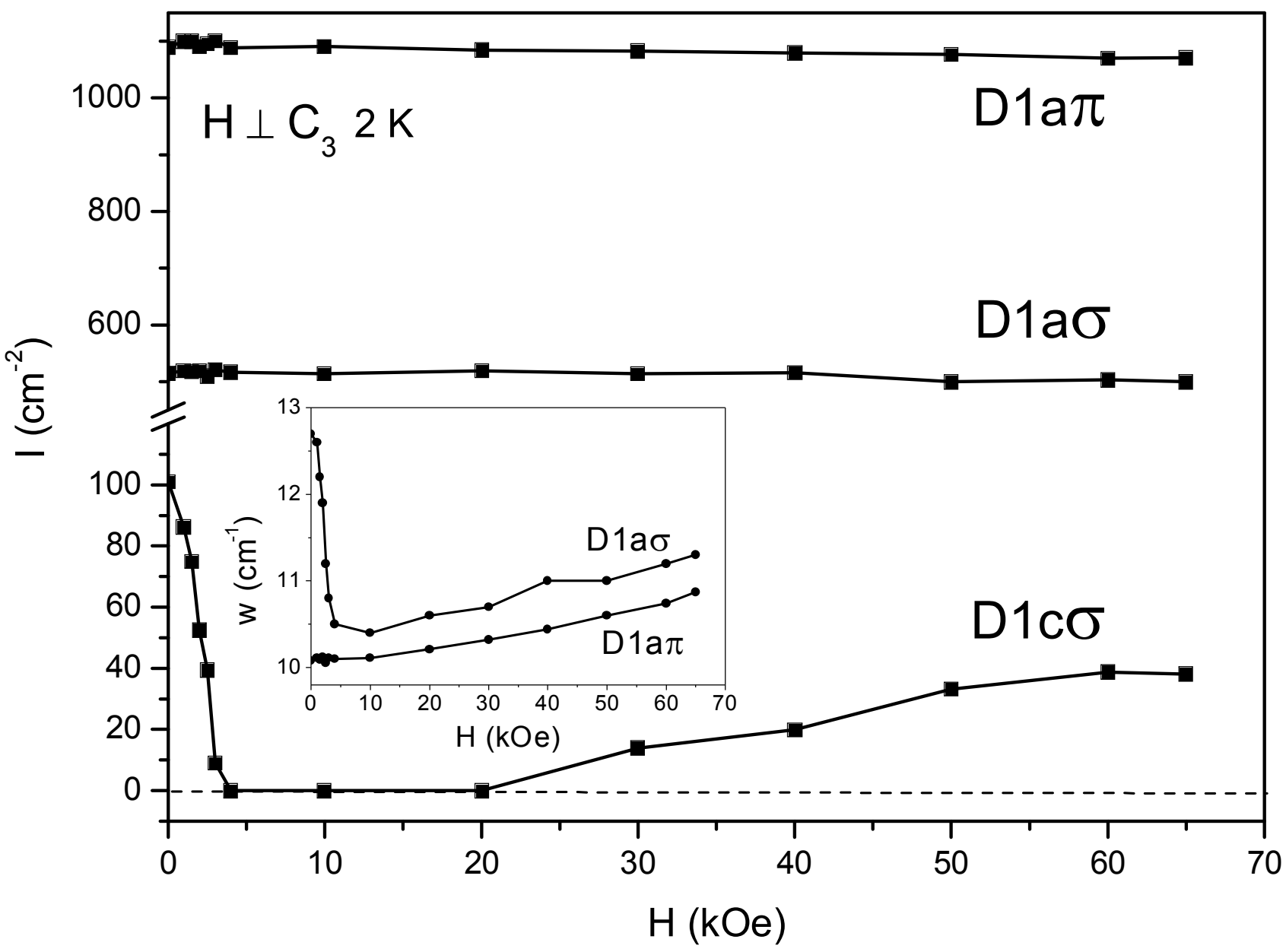

Fig. 6. The D1 transitions intensities as a function of magnetic field H⊥$C_3$ at T=2 K. Inset: line widths as a function of magnetic field H⊥$C_3$.

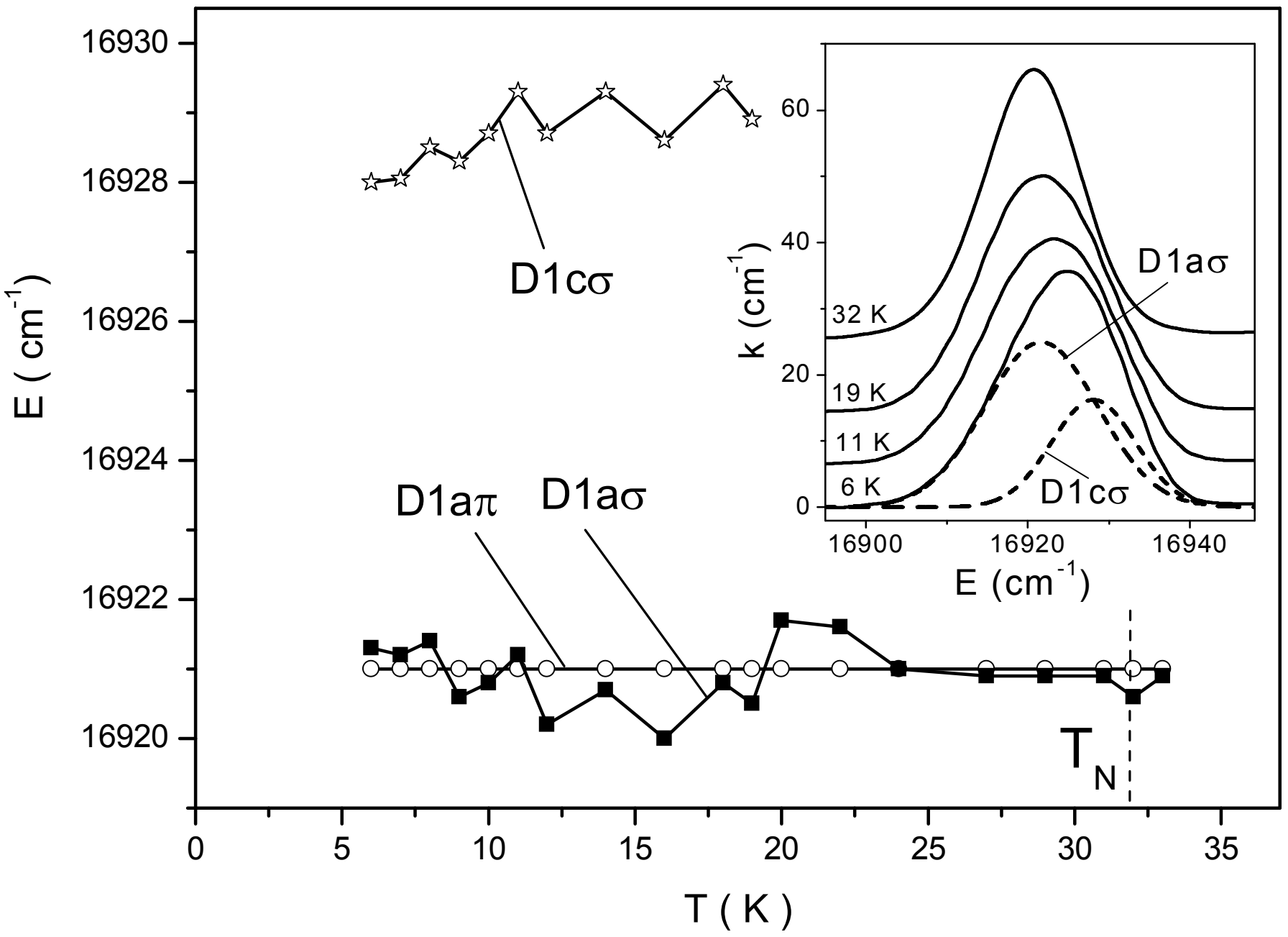

Fig. 7. The D1 transitions energies as a function of temperature in zero magnetic field. Inset: σ-polarized absorption lines at several temperatures.

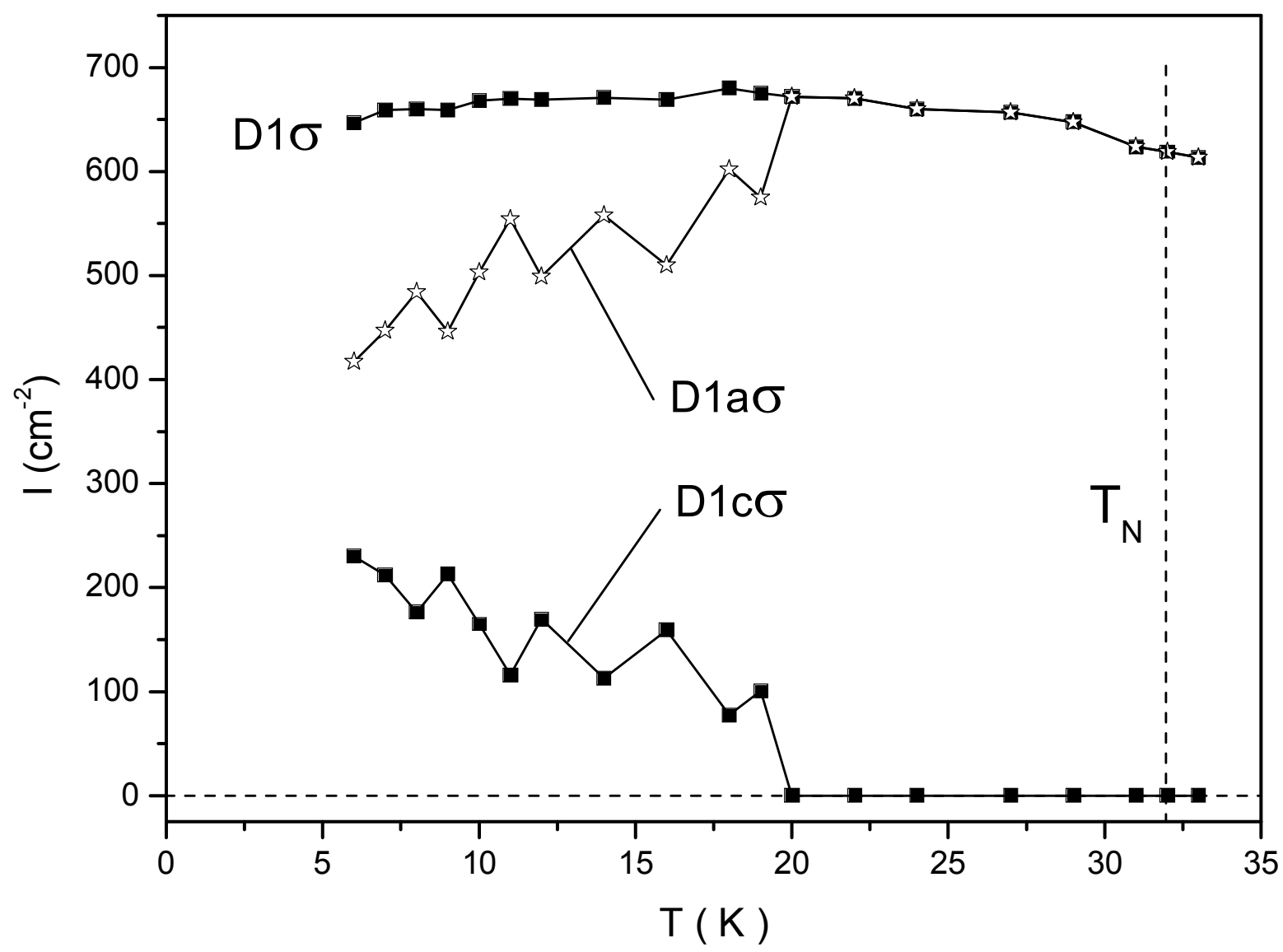

Fig. 8. The D1 transitions intensities as a function of temperature in zero magnetic field.

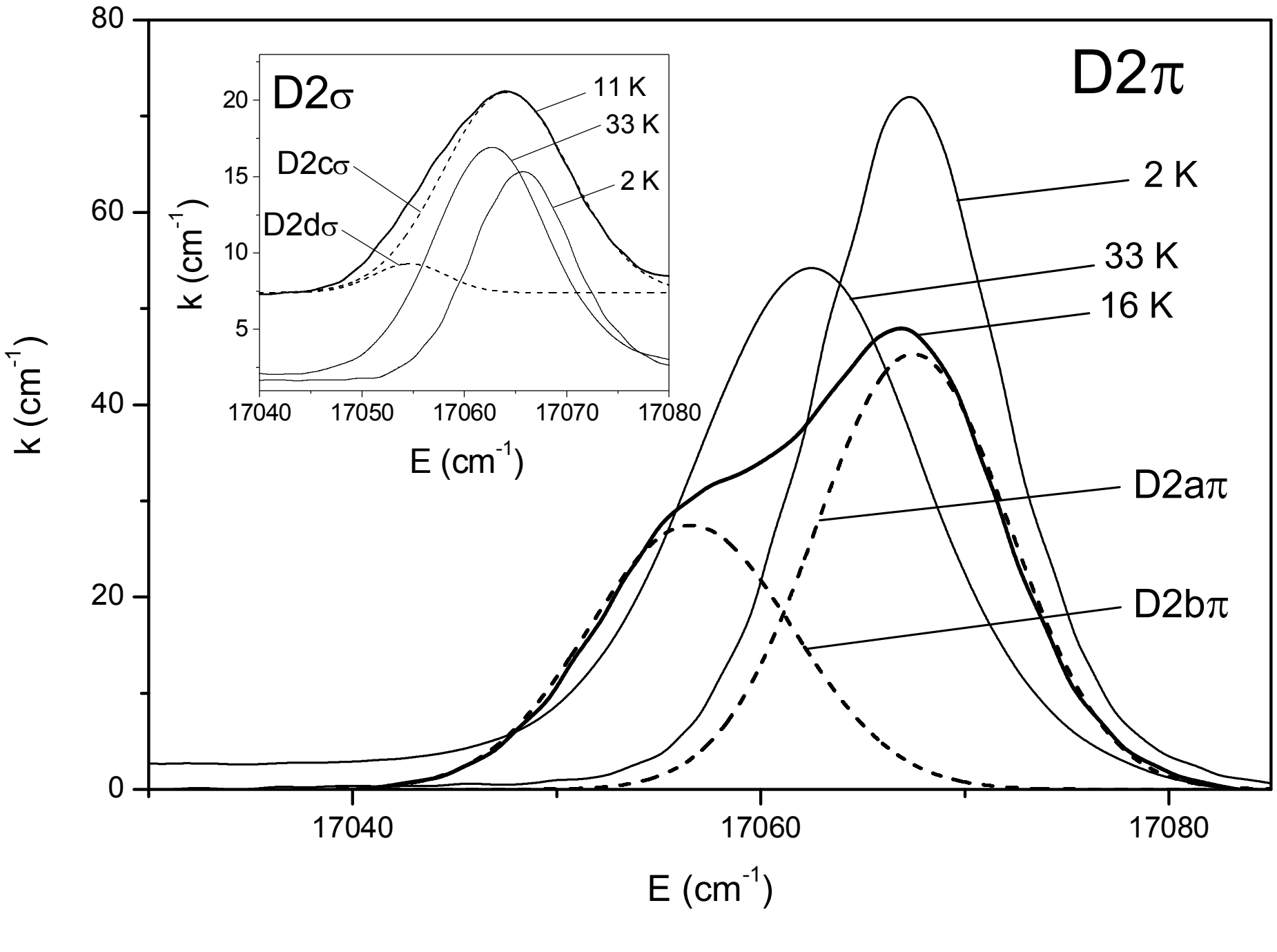

Fig.9. π-polarized absorption spectra of the D2 transitions at several temperatures in zero magnetic field. Inset: the same at σ-polarization.

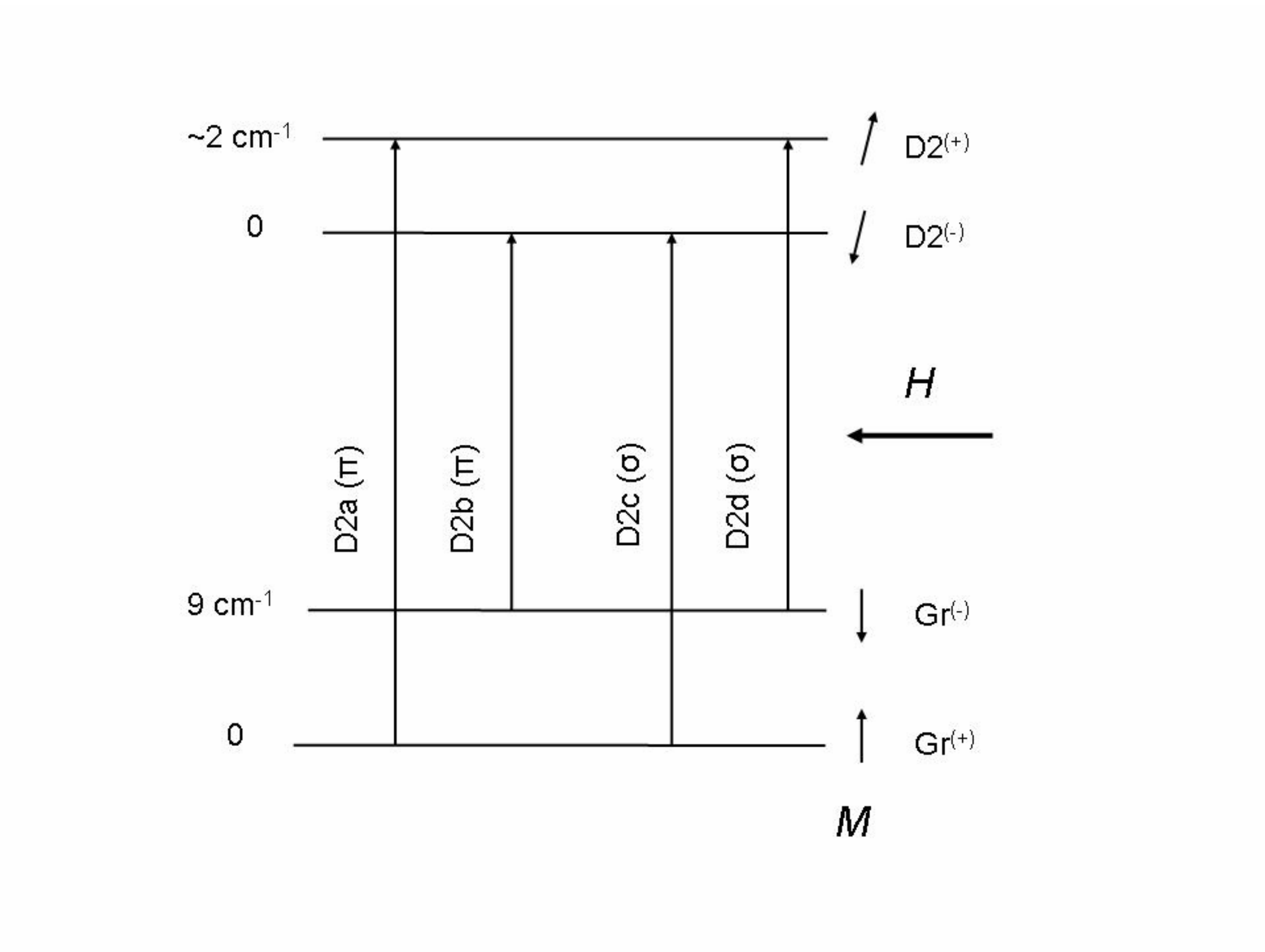

Fig. 10. Diagram of the D2 transitions.

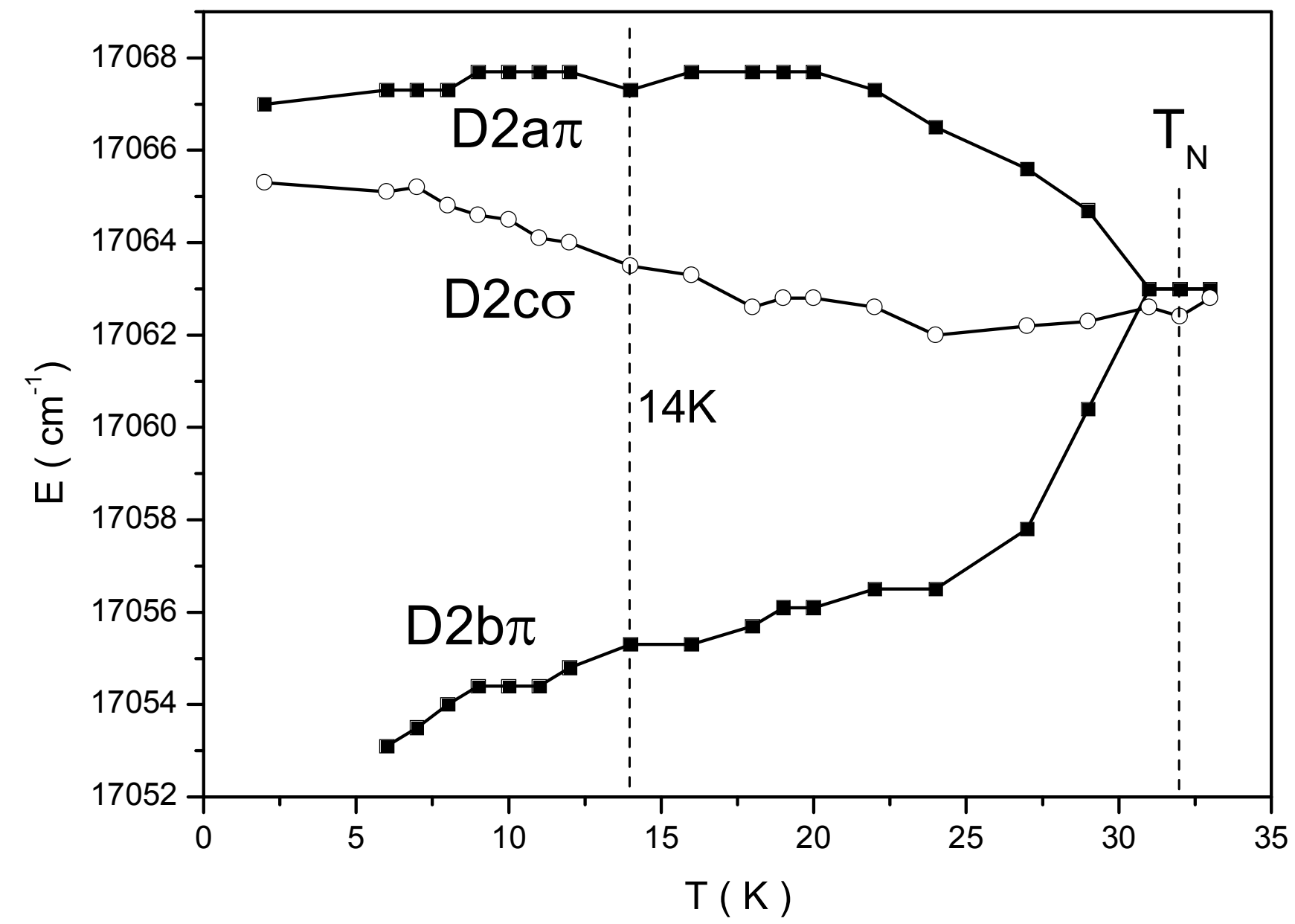

Fig. 11. The D2 transitions energies as a function of temperature in zero magnetic field.

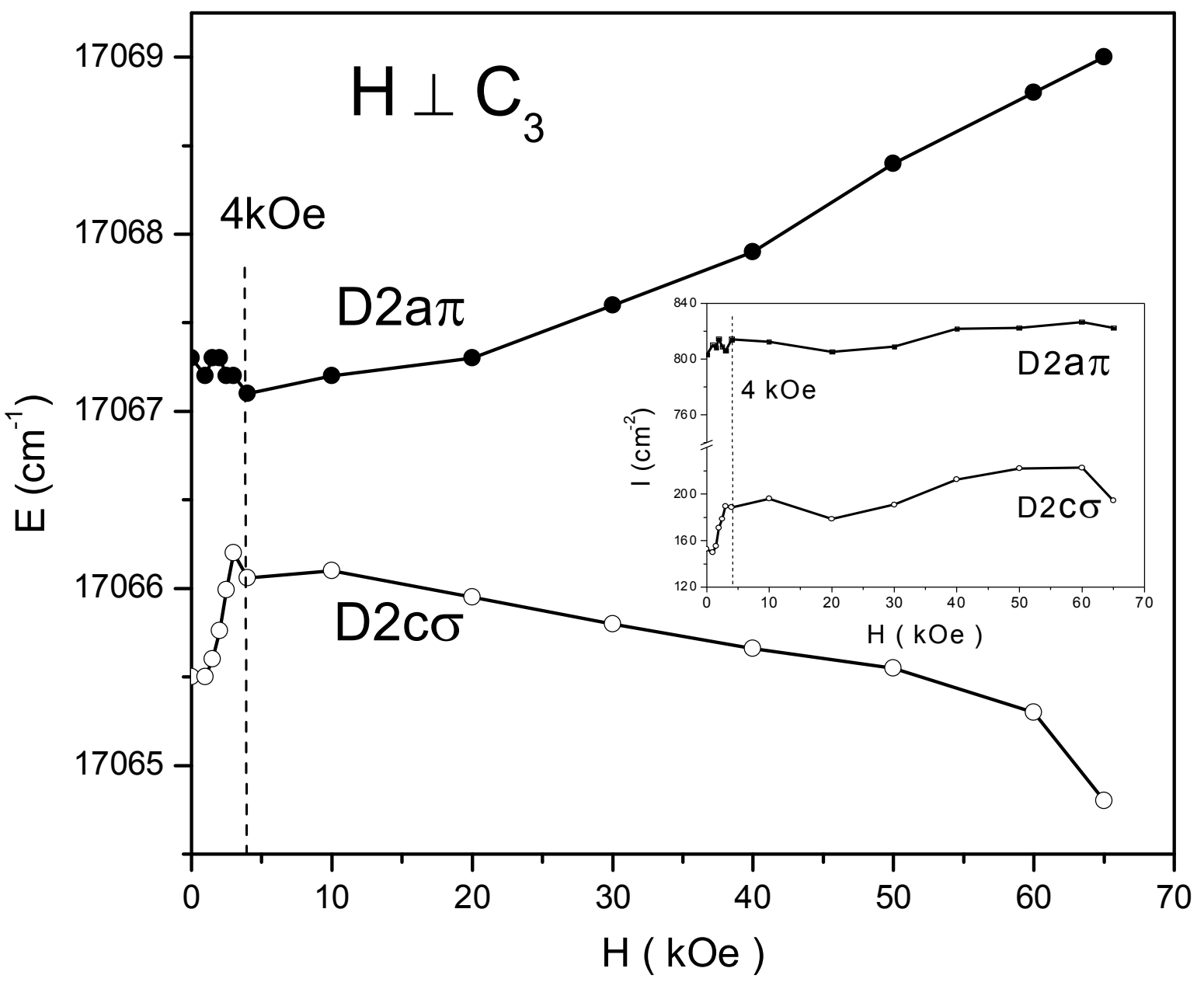

Fig. 12. The D2 transitions energies as a function of magnetic field H⊥$C_3$ at T=2 K. Inset: the same in magnetic field H||$C_3$.

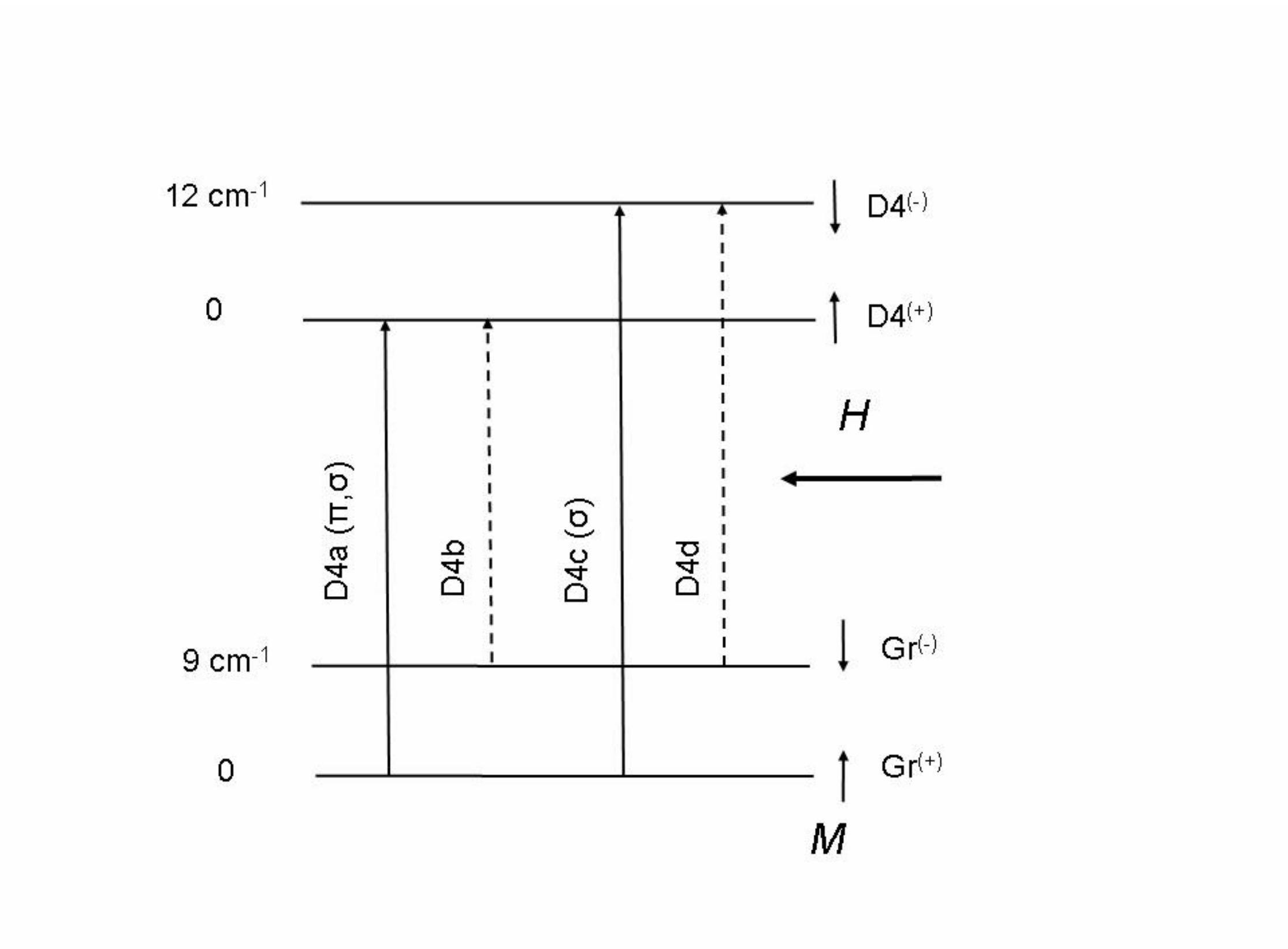

Fig. 13. Diagram of the D4 transitions.

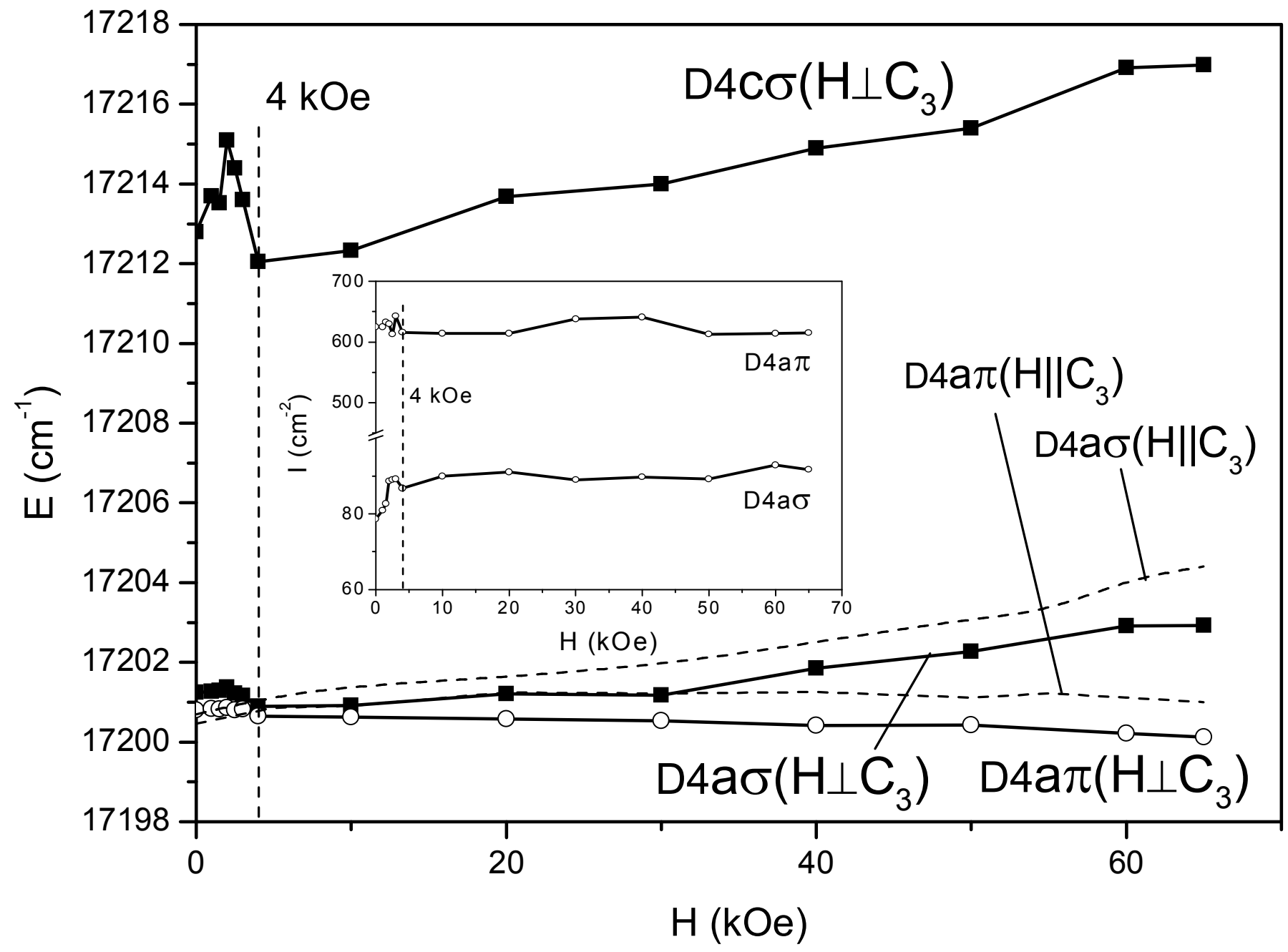

Fig. 14. The D4 transitions energies as a function of magnetic field H⊥$C_3$ and H||$C_3$ at T=2 K. Inset: the D4 transitions intensities as a function of magnetic field H⊥$C_3$ at T=2 K.

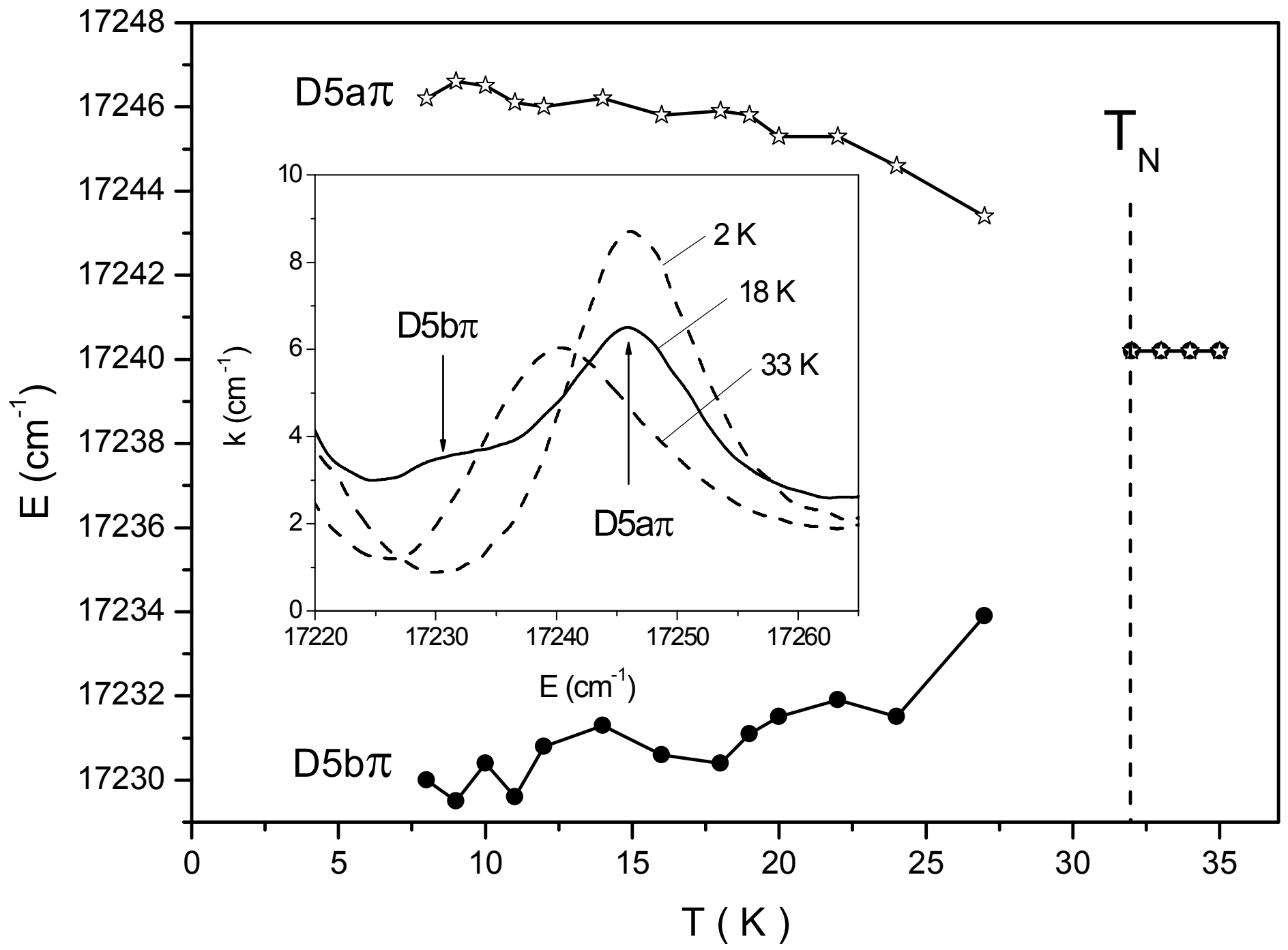

Fig. 15. The D5 transitions energies as a function of temperature in zero magnetic field. Inset: π-polarized absorption lines at several temperatures.

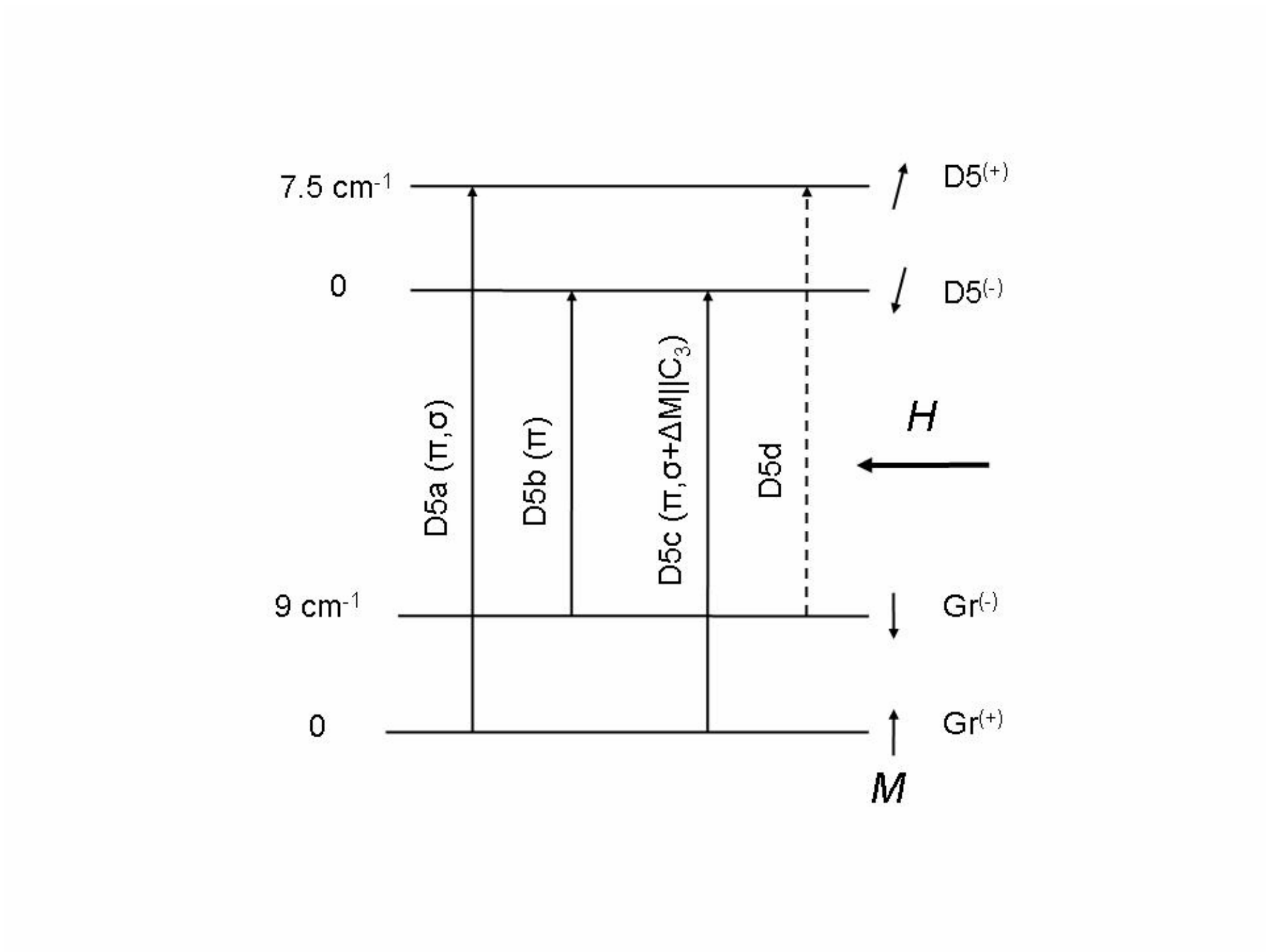

Fig. 16. Diagram of the D5 transitions.

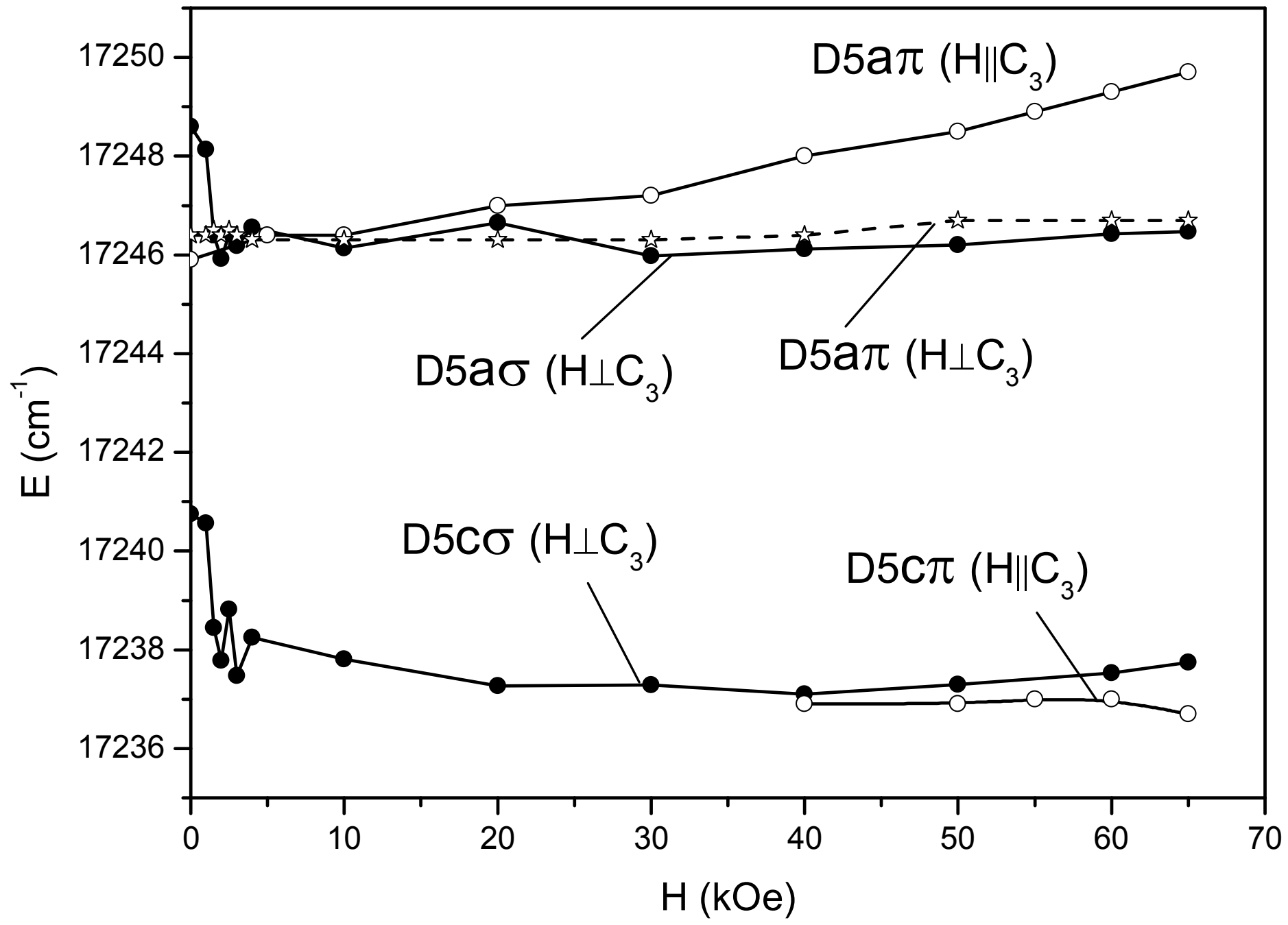

Fig. 17. The D5 transitions energies as a function of magnetic field $H \perp C_3$ and $H || C_3$ at T=2 K.

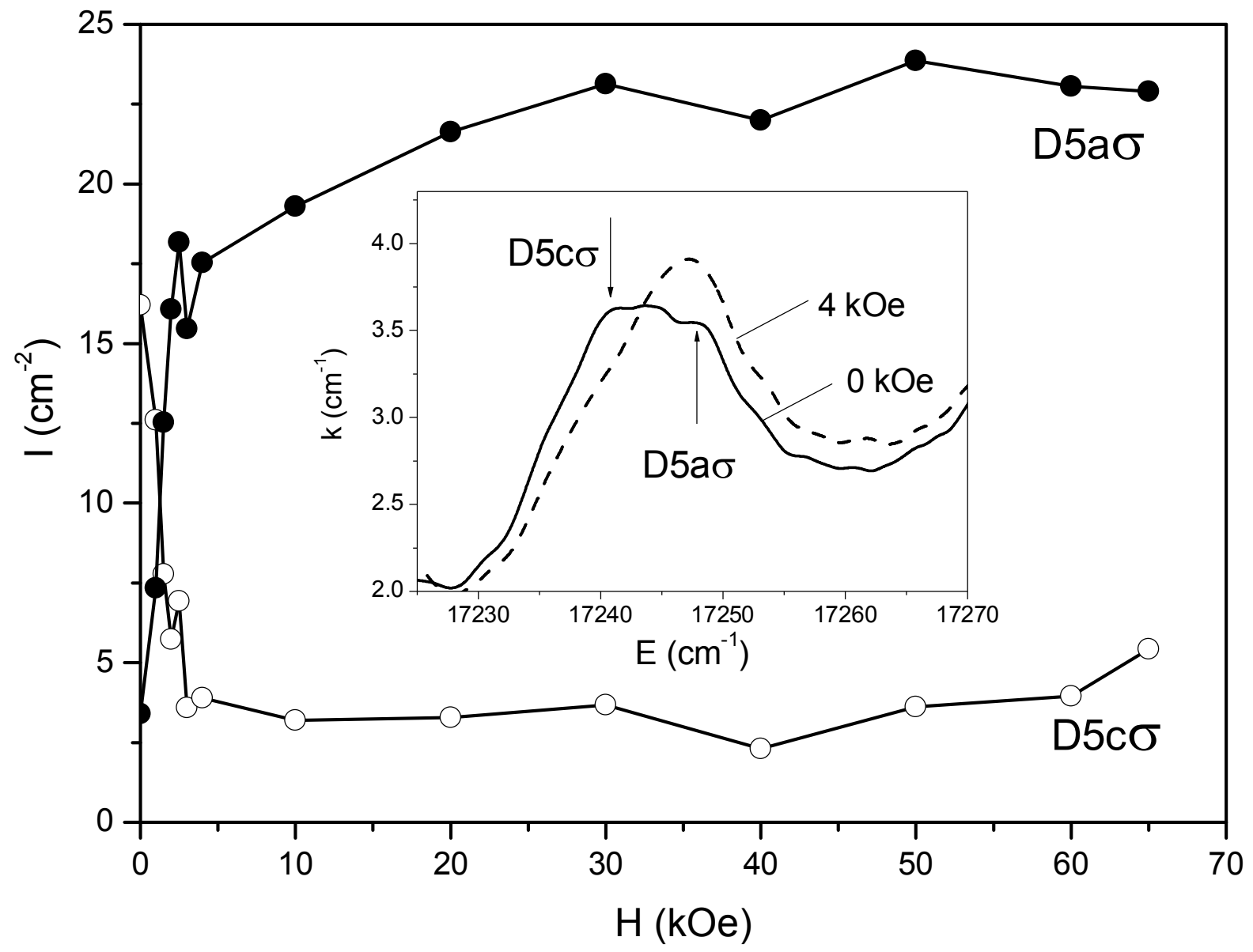

Fig. 18. The D5 transitions intensities as a function of magnetic field H⊥$C_3$ at T=2 K. Inset: σ-polarized absorption spectra of the D5 transition at two magnetic fields.

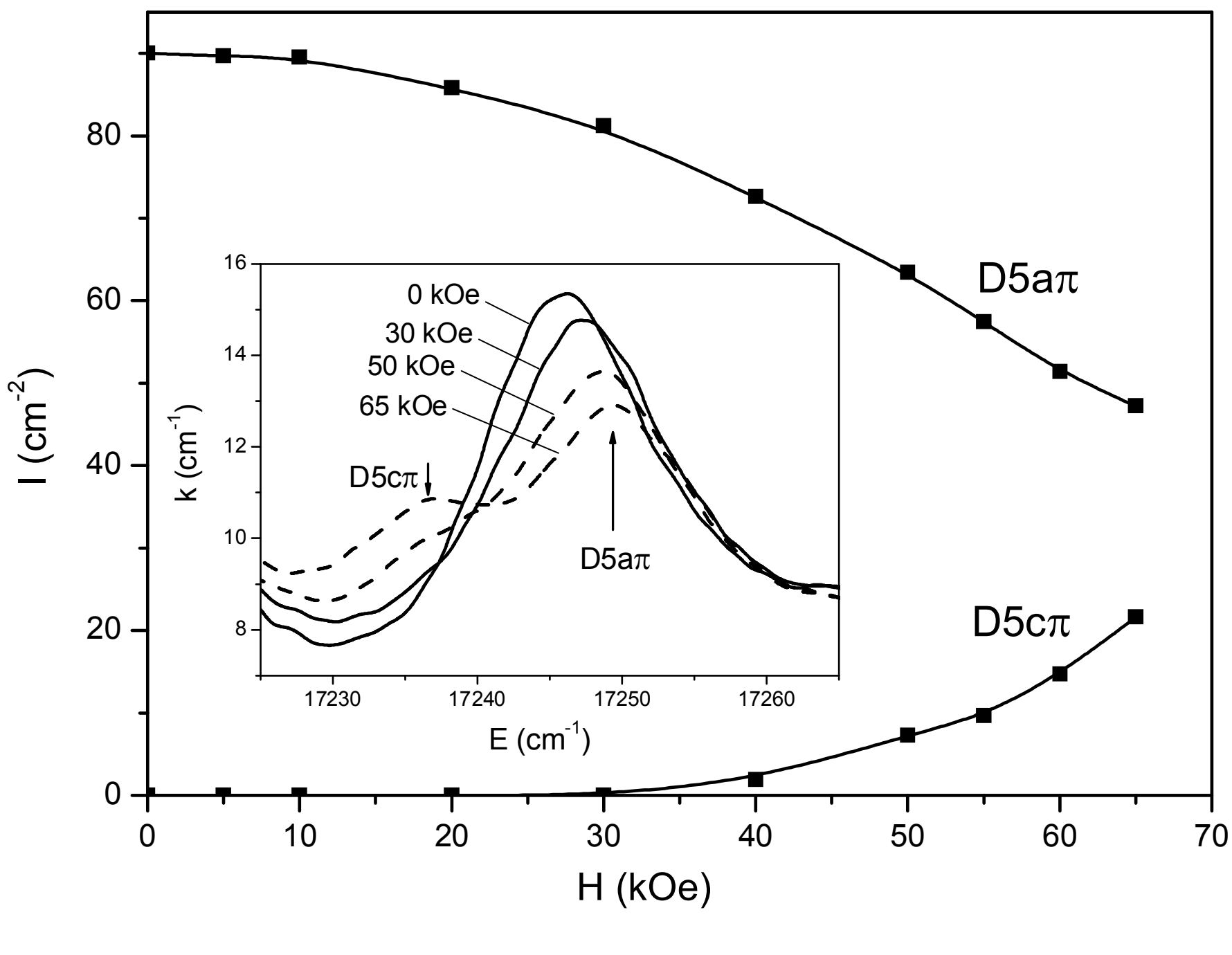

Fig. 19. The D5 transitions intensities as a function of magnetic field H||$C_3$ at T=2 K. Inset: π-polarized absorption spectra of the D5 transition at several magnetic fields.

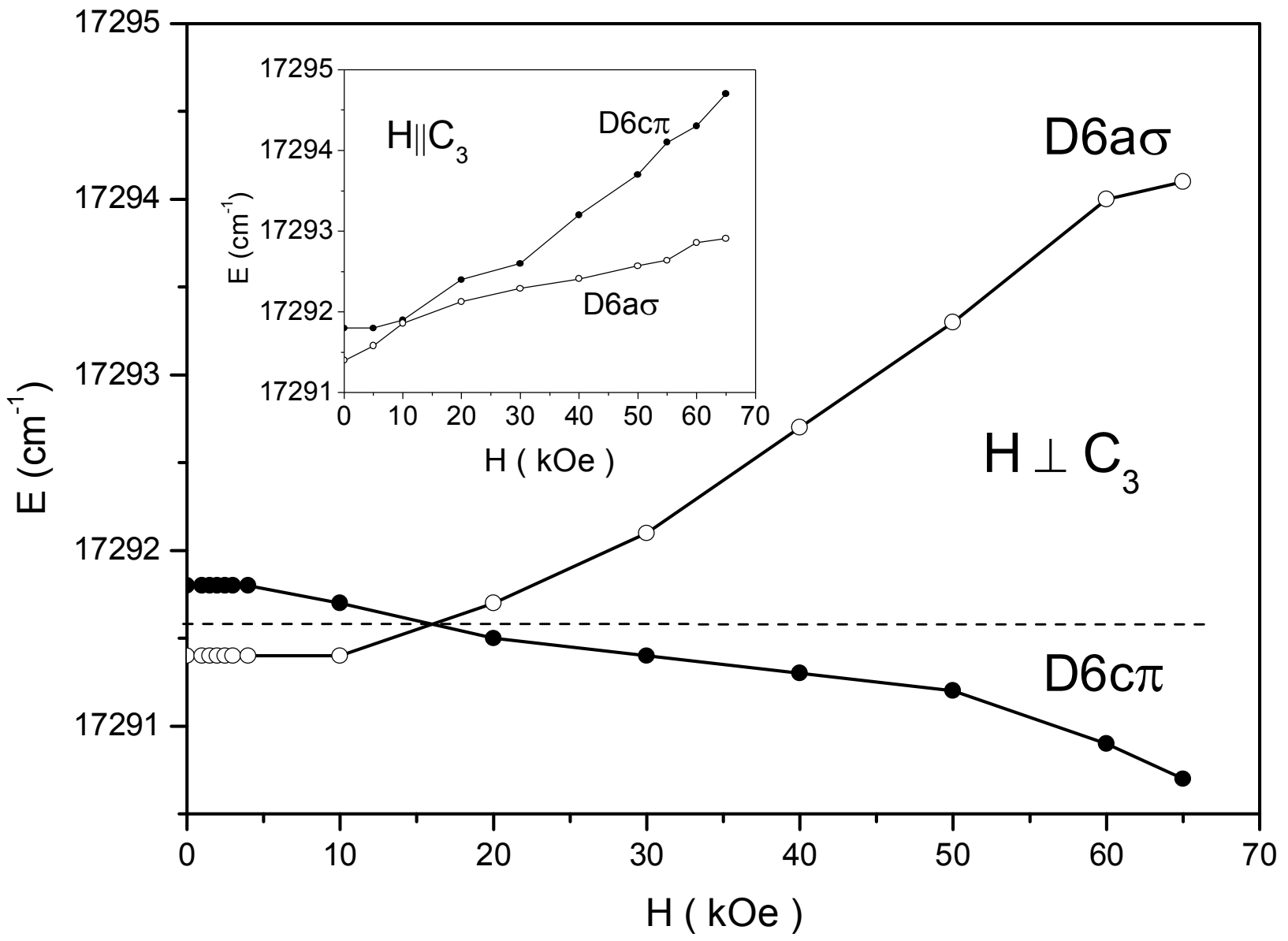

Fig. 20. The D6 transitions energies as a function of magnetic field H⊥$C_3$ and H||$C_3$ at T=2 K.

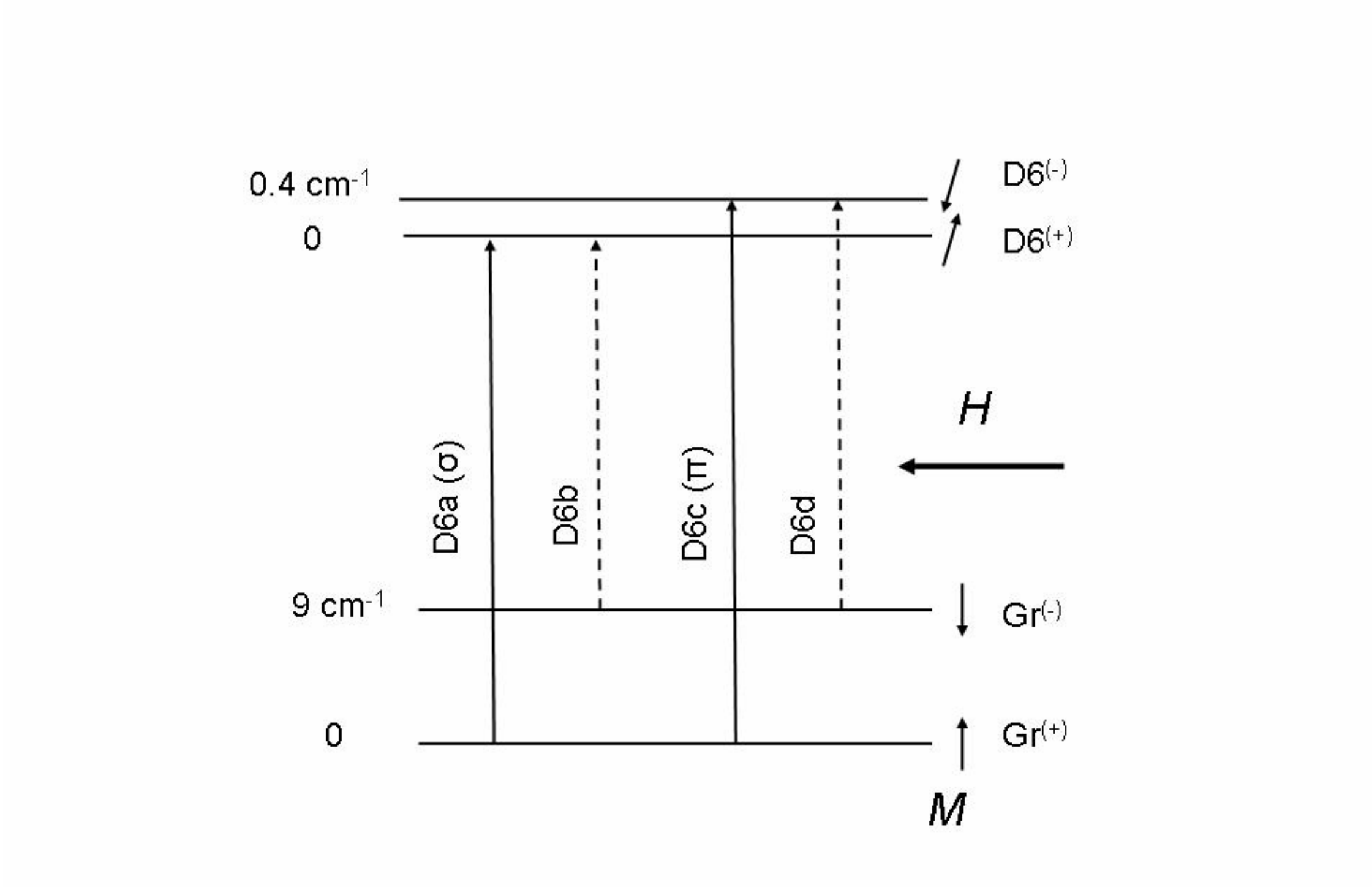

Fig. 21. Diagram of the D6 transitions.

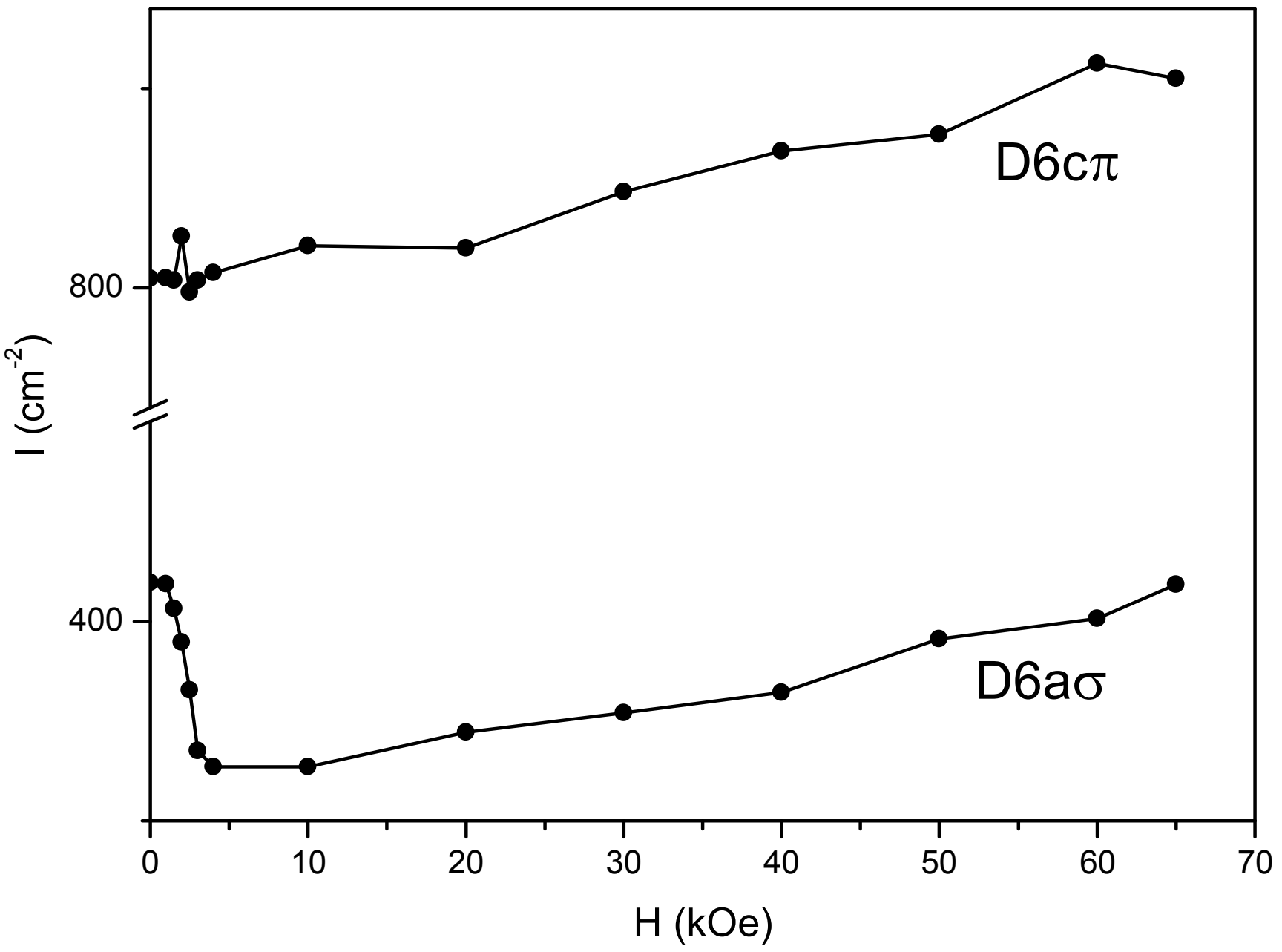

Fig. 22. The D6 transitions intensities as a function of magnetic field $H \perp C_3$ at T=2 K.

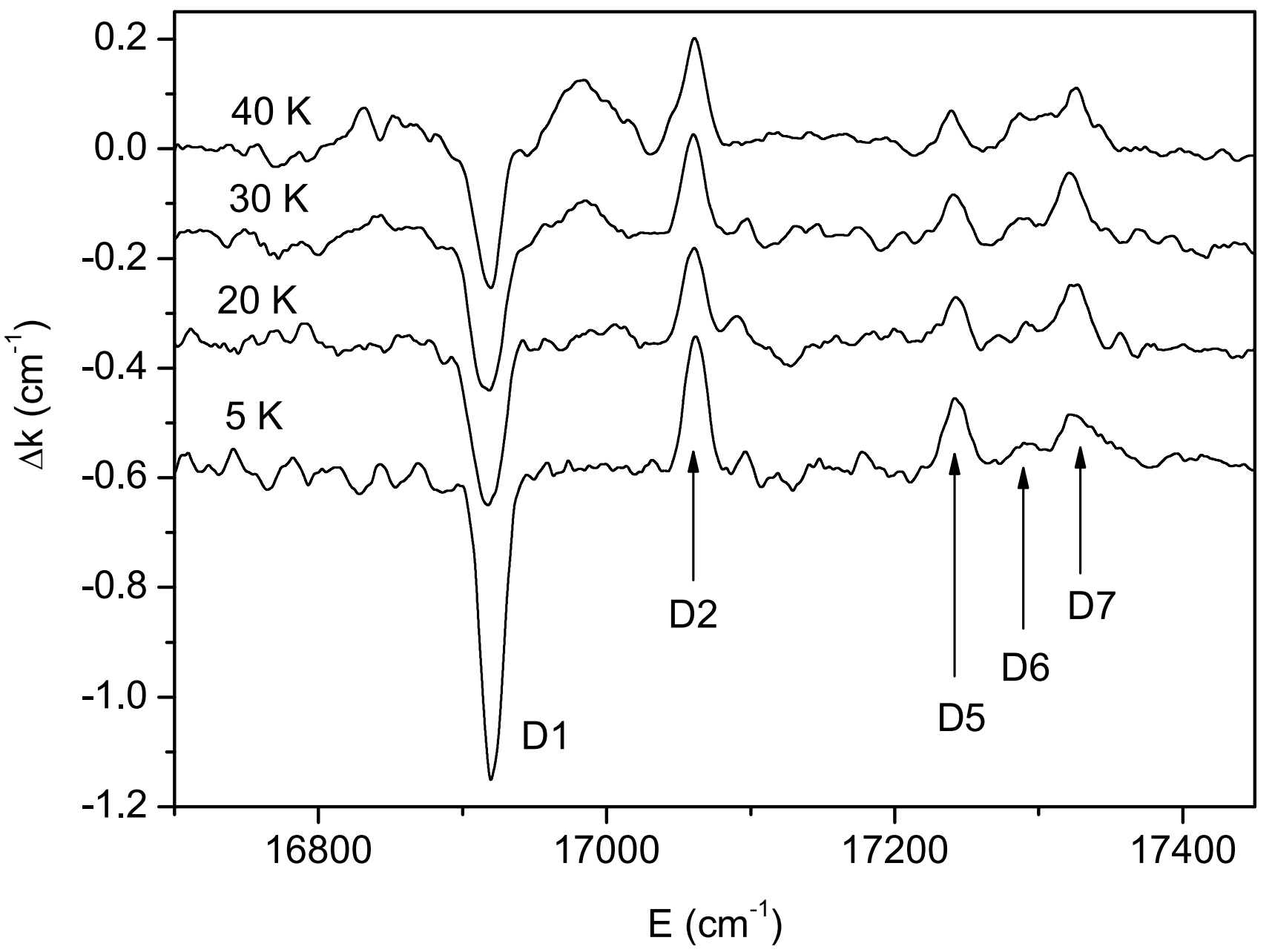

Fig. 23. NCD spectra of the $^4I_{9/2} \rightarrow (^4G_{5/2} + ^2G_{7/2})$ transition (D-band).

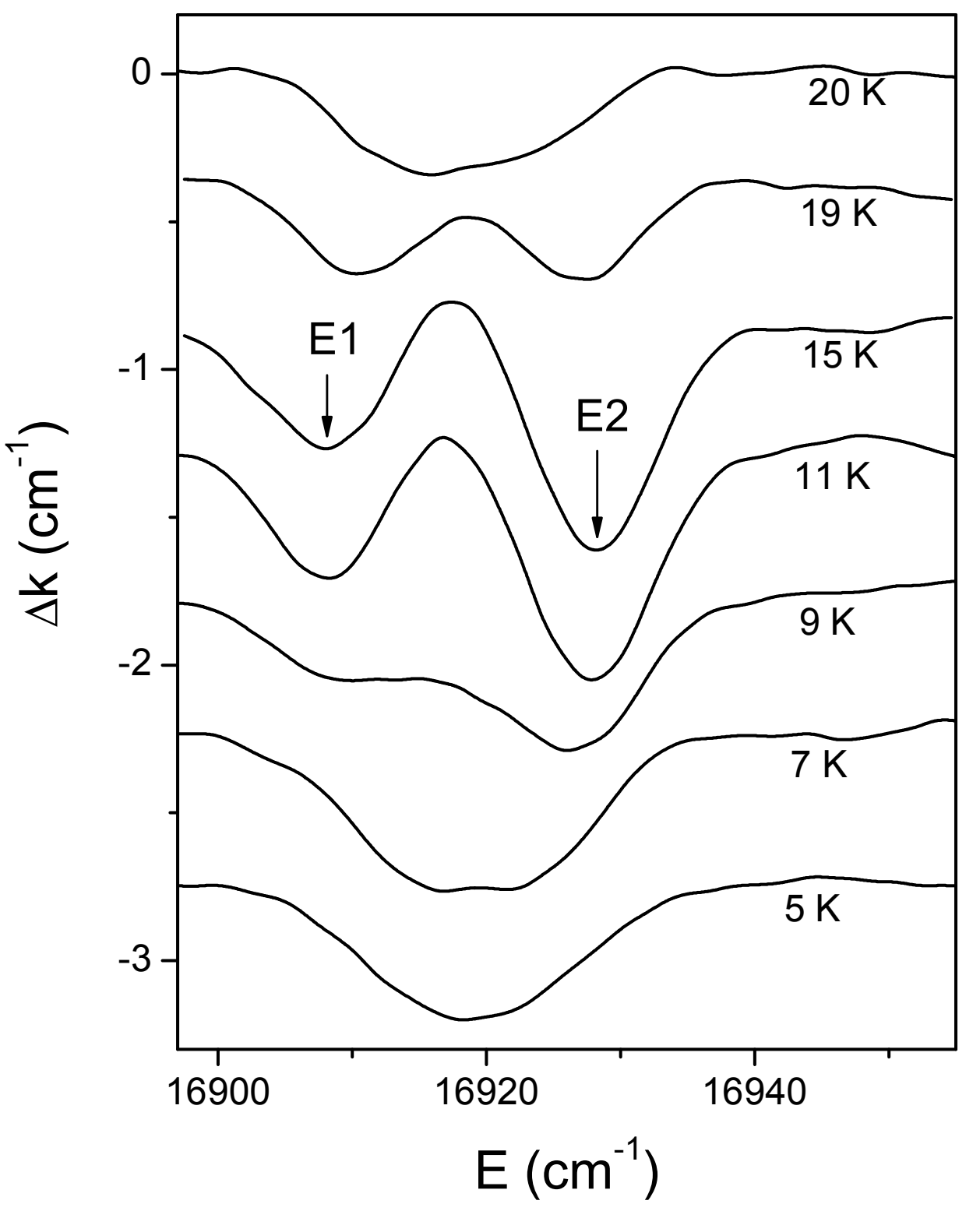

Fig. 24. NCD spectra in the region of the D1 line.

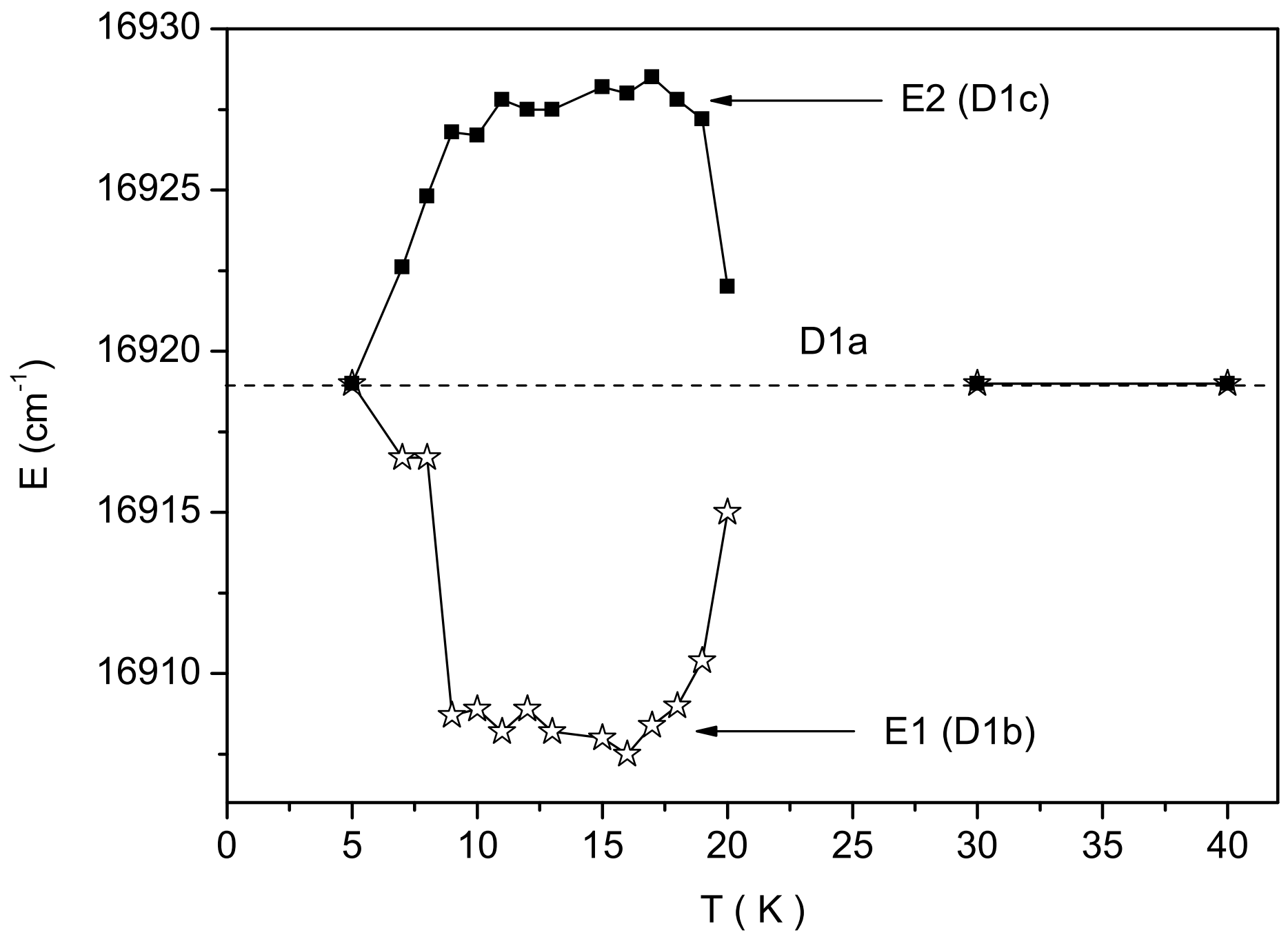

Fig. 25. Energies of components of the NCD spectrum in the region of the D1 line as a function of temperature.